\documentclass[usenatbib,useAMS]{mn2e}
\usepackage{times}
\usepackage{graphicx}
\usepackage{natbib}

\title[Modelling galactic spectra: I - A dynamical model for NGC~3258]{Modelling galactic spectra: I - A dynamical model for NGC~3258\thanks{Based on observations obtained at the European Southtern Observatory, La Silla, Chile (Programmes Nr. 62.N-0492, 64.N-0192)}}
\author[V. De Bruyne et al.]
       {V. De Bruyne$^{1}$, S. De Rijcke$^{1}$\thanks{FWO postdoctoral fellow}, H. Dejonghe$^{1}$\thanks{E-mail:Herwig.Dejonghe@rug.ac.be}, W.W. Zeilinger$^{2}$ \\
        $^{1}$Astronomical Observatory, Ghent University, Krijgslaan 281, S9, 9000 Ghent, Belgium\\
$^{2}$ Institut f\"ur Astronomie, Universit\"at Wien, T\"urkenschanzstrasse 17, A-1180 Wien, Austria
}
\date{Accepted 
      Received ;
      in original form }

\pagerange{\pageref{firstpage}--\pageref{lastpage}}
\pubyear{}

\begin{document}

\maketitle

\label{firstpage}

\begin{abstract}

In this paper we present a method to analyse absorption line spectra
of a galaxy designed to determine the stellar dynamics and the stellar
populations by a direct fit to the spectra. This paper is the first
one to report on the application of the method to data.  The modelling
results in the knowledge of distribution functions that are sums of
basis functions. The practical implementation of the method is
discussed and a new type of basis functions is introduced.

With this method, a dynamical model for NGC~3258 is constructed. This
galaxy can be successfully modelled with a potential containing 30\%
dark matter within $1r_e$ with a mass of $1.6\times 10 ^{11}
M_\odot$. The total mass within $2r_e$ is estimated as $5\times10^{11}
M_\odot$, containing 63\% dark matter. The model is isotropic in the
centre, is radially anisotropic between 0.2 and 2 kpc ($0.88 r_e$) and
becomes tangentially anisotropic further on.  The photometry reveals
the presence of a dust disk near the centre.
\end{abstract}


\section{Introduction}

It is well known that in galaxies stellar populations and stellar
dynamics cannot be studied completely independently; the determination
of the line profiles and kinematic parameters depends on assumptions
about the dominant stellar population, while the determination of the
chemical properties depends on the absorption strenght of the spectral
lines.

In general, stellar kinematics are retrieved by comparing the observed
spectra with the convolution of a stellar template spectrum with a
model line-of-sight velocity profile, see \citet{db2} for a discussion
of some standard techniques. The stellar template belongs to the same
spectral class as the dominant stellar population in the galaxy, or
can be a mix of different stellar type spectra. Often, a best matching
stellar mix is determined for the spectrum at the centre, and this mix
is then used throughout the whole galaxy.  In some cases, the stellar
mix is allowed to vary with radius, (e.g. \citet{cbh}). However, the
only criterion used for obtaining a proper mix is a good
fit, without any physical considerations. Therefore, it is
acknowledged that this procedure does generally not yield a reliable
population synthesis.

The kinematics obtained from a parametrization of the line-of-sight
velocity distributions can be used as input for a dynamical model.
Within this multistep scheme of dynamical modelling, it is
acknowledged that dynamical evidence for dark matter in elliptical
galaxies requires the use of more information contained in the
line-of-sight velocity distribution (hereafter LOSVD) than simply the
mean velocity and velocity dispersion. The reason for this is the
existence of a so called mass-anisotropy degeneracy. In a dynamical
model it is possible to mimic the influence of a dark matter halo by
tangential anisotropy. Therefore, also the anisotropy information in
LOSVDs should be used. However, there is no uniformity in the
parametrization of these LOSVDs and different algorithms may well have
different biases \citep{jos}.  Moreover, it is not always
straightforward to include these parameters in a fit, since not all of
them are linearly dependent on the distribution function (hereafter DF)
(\citet{gerh}, \citet{db}).  Another option is to use the complete
LOSVD as dynamical input. This omits the need to approximate the LOSVD
with a number of parameters, but in combination with the Schwarzschild
orbit superposition method, this requires a careful smoothing both of
the LOSVD and the dynamical model \citep{gebh}. But there is still
another way of dealing with the kinematic information of a galaxy,
through direct modelling of the observed spectra.

In this paper we adopt a method to analyse spectra from elliptical
galaxies that explores the dynamics of the galaxy and simultaneously
offers a way to study the stellar populations in that galaxy.  The
method originates from the field of dynamical modelling and is
outlined by \cite{dr}, hereafter DD98.  The main idea is that the different stellar
populations that contribute to the integrated galaxy spectrum do not
necessarily share the same kinematic characteristics. Hence, in terms
of dynamical modelling, they should be associated with different
distribution functions.  A simultaneous determination of the best
stellar mix and the best distribution functions associated with them
de facto amounts to a dynamical model and a stellar population
synthesis.

This paper is the first to present the application of the modelling
method to data. The aim is to determine the gravitational potential
of NGC~3258, by modelling the galaxy spectra using different stellar
templates.  The modelling strategy is outlined and commented, and a
new family of dynamical components is presented.

This is the first paper in a series of two, where a deeper interplay
between dynamics and population synthesis is explored. In this first
paper, the attention will be focused on the dynamical modelling
aspect, while the second paper will go deeper into the stellar
population analysis.

In the next section, the observations and data reduction steps are
presented.  The modelling strategy is outlined in section 3.  Section
4 presents the dynamical components that were used. In section 5 the
results of the modelling are presented. Section 6 contains a
discussion of the results and the conclusions can be found in section
7.

\section[]{Observations and data reduction}

The E1 galaxy NGC~3258 is one of the brightest members of the Antlia
Group. In its neighborhood, NGC~3257 can be found at 4.5 arcmin (about
$21 r_e$) and NGC~3260 is at 2.6 arcmin (about $12 r_e$). We measured
a systematic velocity of 2709 km/s, this puts NGC~3258 at a distance
of 36.12 Mpc (taking $H_0=75 \rm{km/s/Mpc}$). This means that a linear
size of 1 kpc at the distance of NGC~3258 translates into an angular
size of $5.7''$.

NGC~3258 was observed with the ESO-NTT telescope in the nights of
27-28/2/2000. Spectra of the major axis covering the Ca II triplet
around 8600 {\AA} were taken. During the observations in 2000, also
imaging of NGC~3258 in the R band was performed. The detector used for
this observation run was a Tektronix CCD with $2048\times 2047$
pixels, 24$\mu m \times$24$\mu m$ in size and with a pixel scale of
$0.27''$/pix. Another observation run, in 1999, was used to obtain
images in the B band.

\subsection{Photometry}
R band observations (in total 600 sec integration time) of NGC~3258
were obtained at the ESO-NTT using the red arm of EMMI during the
spectroscopic run in februari 2000. An additional B band image (600
sec integration time in total) was obtained in february 1999 in the
same configuration. 
The detector used for
the B band images was a Tektronix CCD with 1024$\times$1024 pixels,
24$\mu m \times$24$\mu m$ in size and with a pixel scale of
$0.37''$/pix.

\begin{figure}
\includegraphics[bb=20 -70 470 720,scale=.5,clip=true]{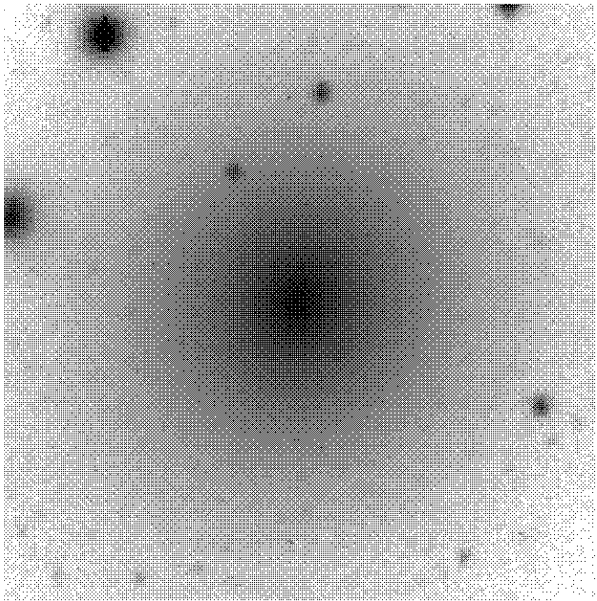}
  \caption{B band image of NGC 3258, the size of the region is $110''\times110''$. North is up and east is to the left. }
 \label{ngc3258}
\end{figure}

\begin{figure}
\includegraphics[bb= 120 40 490 410,scale=.55,clip=true]{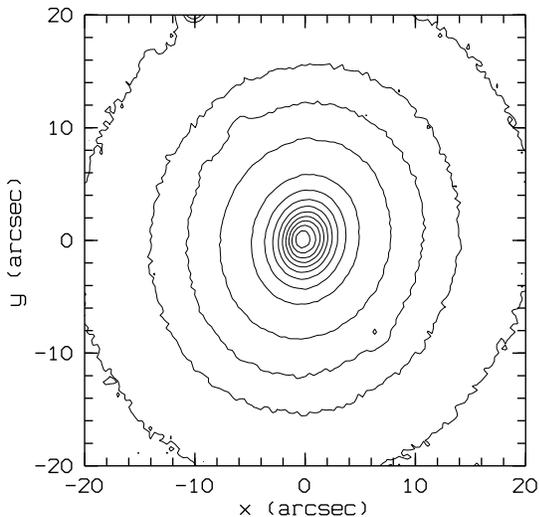}
  \caption{Isophotes of NGC~3258.}
  \label{contour}
\end{figure}

The seeing (judged from a number of stars in the images) was $1.1''$
for the R band images and $1.7''$ for the B band images. The
photometric zero point was established by comparison with published
photometric data available on NED.  Figure \ref{ngc3258} shows an
image of NGC~3258, isophotes are plotted in figure \ref{contour}.  In
this figure, the central isophote seems to be a little displaced
compared to the other isophotes.
Both B and R images show a slightly off-centre nucleus, with a shift
of about $0.45''$ (or 0.07 kpc) in the horizontal direction and a
shift of about $0.1''$ (or 0.01 kpc) in the vertical direction. This
is illustrated in the upper panel figure \ref{dust2}, where the shift
of the centres of the isophotes is shown.

\begin{figure}
\includegraphics[bb=30 245 460
  685,clip=true,scale=.55]{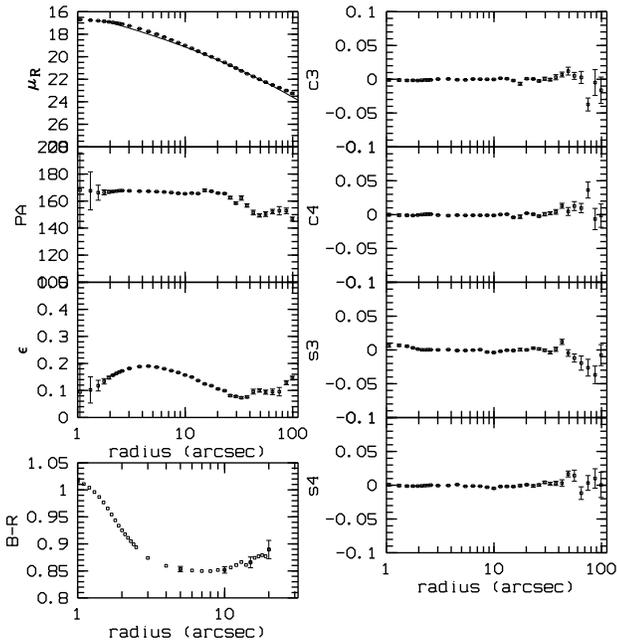} \caption{Photometric
  parameters for NGC~3258. Left column, from top to bottom: surface
  brightness, position angle, ellipticity and B-R. Right column, the
  coefficients indicating deviations from pure elliptic isophotes.}
  \label{fotometrie}
\end{figure}

The photometric parameters were derived by fitting elliptic isophotes
to the image. The intensity along the isophotes was expanded in a
fourth order Fourier series, in order to reveal deviations from pure
elliptic isophotes, like in e.g. \citet{rijdi}.  In figure
\ref{fotometrie}, the main photometric parameters are shown: surface
brightness, position angle, ellipticity and B-R profile as derived
from the B and R surface brightness profiles in the left column.  The
coefficients indicating deviations from pure elliptic isophotes are
shown in the right column. The values for these coefficients indicate
that NGC~3258 is an almost perfect elliptical galaxy, with an
ellipticity between 0.1 and 0.2.

A S\'ersic profile ($I(r) = I_0 \exp[-(r/r_0)^{1/n}]$, with $I(r)$ the
surface brightness profile, $I_0$ the central surface brightness,
$r_0$ the scale length, an $n$ the S\'ersic parameter \citep{ci} was
fitted to the surface brightness, the result of the fit is overplotted
(in solid line) in the upper left panel of figure \ref{fotometrie}. The
observed surface brightness profile can be well matched with a
S\'ersic profile with S\'ersic parameter $n=5.47\pm 0.07$.  The
effective radius is $12.95''$ or 2.26 kpc.

Values for B-R (derived from the photometric profiles) are shown in
the lower left panel of figure \ref{fotometrie}.  Within the central
$5''$, there seems to be a slight reddening of the nucleus. B-R
changes from about 1 in the most central measurement to about 0.85 at
$5''$.

\subsection{Central dust component}
\begin{figure}
\includegraphics[bb=130 240 375 450,clip=true,scale=1]{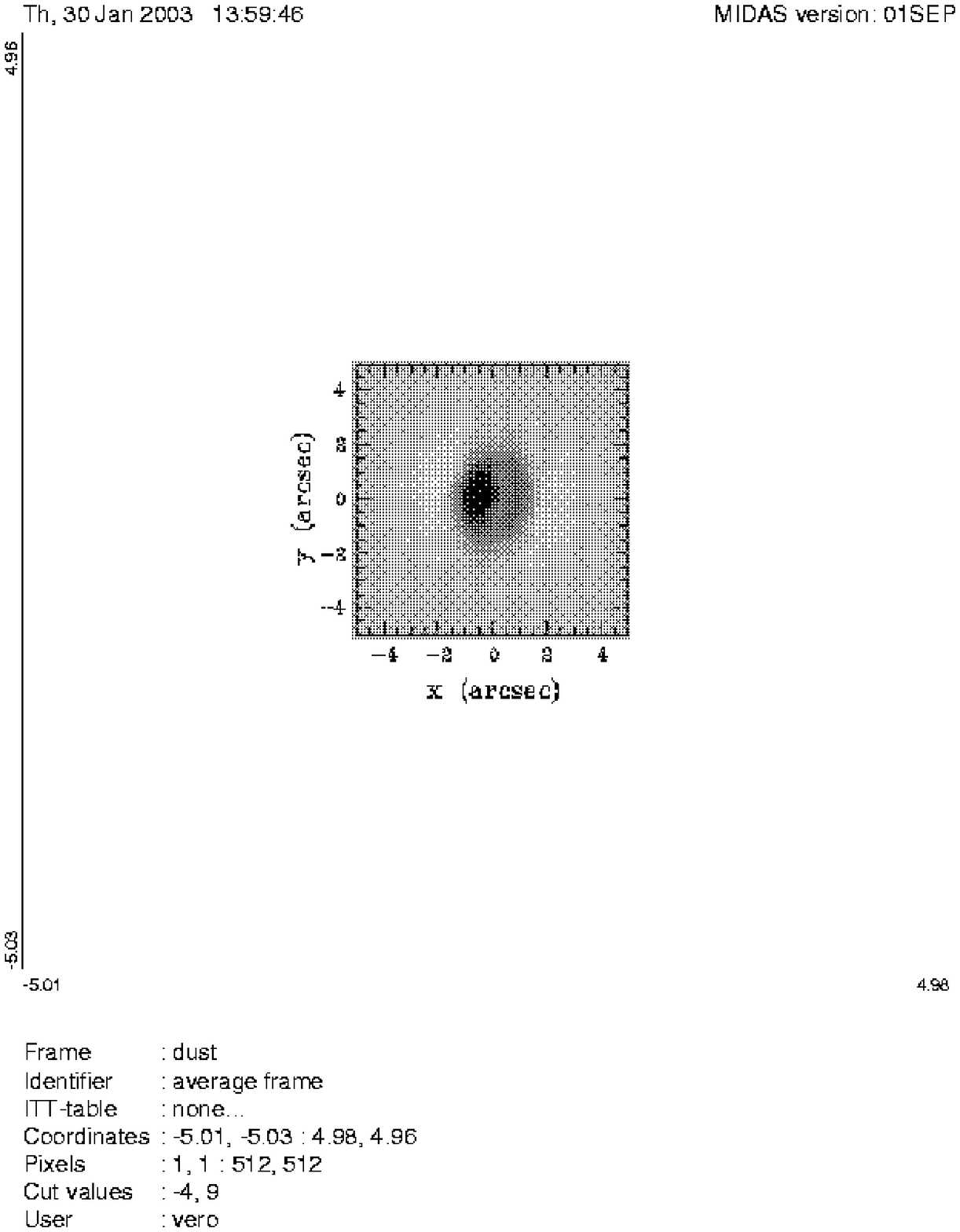}
\caption{Inverted image of the central part of the unsharp masked R
image. The light grey band in the centre is a central dust disk.}
\label{dust}
\end{figure}
\begin{figure}
\includegraphics[bb=40 50 395 740,clip=true,scale=.55]{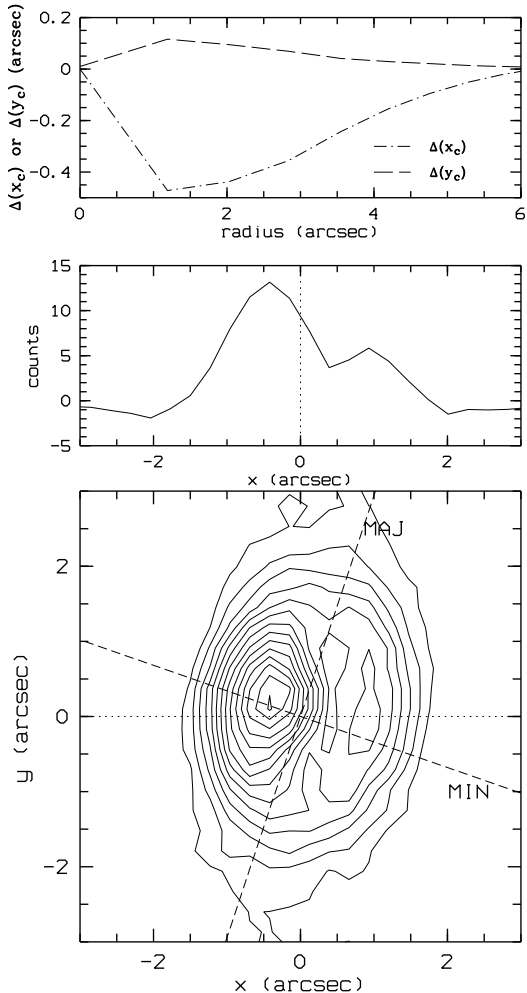}
\caption{ Upper panel: displacement in x ($\Delta(x_c)$) and y
($\Delta(y_c)$) direction of the centre of the isophotes as function
of the major axis of the isophotes. Middle panel: The intensity
profile along the cut with a horizontal axis through the galaxy centre
(dotted line in lower panel). Lower panel: contours of the central
part of the unsharp masked R image. The photometric major and minor
axis are plotted in dashed line.}
\label{dust2}
\end{figure}

Figure \ref{dust} shows the central part of the unsharp masked R image
with a clearly visible dust feature. The estimated radius of the major
axis of the disklike feature is about $1''$ - $1.5''$ (0.17 kpc -
0.26 kpc). Figure \ref{dust2} shows a contour map of this region
(lower panel) and a cut along a horizontal line through the centre
(upper panel). This line does not coincide with any of the photometric
axes. The effect of the dust lane on the apparent brightness is
clearly visible in the upper panel of that figure. The brightest
central point does not coincide with the centre of the isophotes for
the main body of the galaxy. The same behaviour is seen in the B band
images. The off-centre nucleus appearing in the B and R images is
probably due to the obscuring effect of the dust lane.

\subsection{Spectroscopy}
For the spectra, grating \#7 was used, having a dispersion of 0.66
{\AA}/pix.  
A slit width of $1.5''$ yielded a spectral
resolution of 3.67 {\AA} FWHM, resulting in an instrumental dispersion
of about 54 km/s in the region of the Ca II triplet.  For NGC~3258,
several exposures of 3600 sec were taken (in total 11 hours).

A number of standard stars (G dwarfs and K and M giants) were also
observed, they are listed in table \ref{temps}.

Standard reduction steps were applied to these spectra with ESO-MIDAS
\footnote{ESO-MIDAS is developed and maintained by the European Southern
Observatory}.  Dedicated exposures, taken during both runs, were used
for bias correction, dark correction and flatfielding.  Cosmic ray
hits were removed using a top hat filter. After this correction,
remaining cosmics were also removed by hand.

For the wavelength calibration, lamp spectra were taken just before or
after each of the spectroscopic observations of the galaxy or the
stellar template stars. For each row in the spectra a polynomial was
constructed to transform the pixel scale into wavelength scale. The
calibrated images with the Ca II triplet have a step of $0.3${\AA}.

Airmass correction is applied, using the mean value of the airmass at
the beginning and end of the exposure.  The contribution of the sky to
the spectra was estimated from an upper and lower region of the image,
where there was no contribution of the light of the galaxy or template
star. Several galaxy spectra were reduced separately, aligned and
combined into one galaxy spectral image.

\subsection[]{Kinematic parameters}\label{data}

\begin{figure}
\includegraphics[bb=30 150 520 575,clip=,scale=.4]{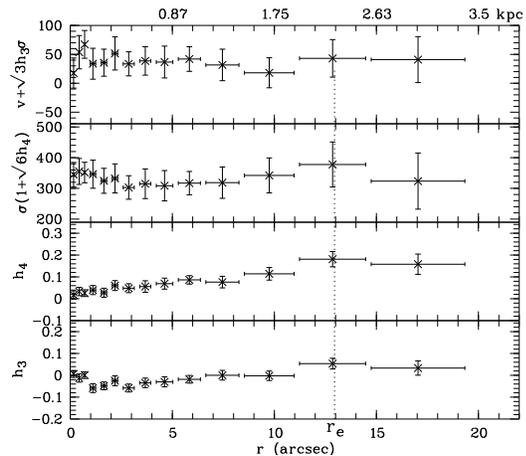} \caption{Kinematics obtained from
  the spectral feature around 8600 {\AA}.}
\label{kin}
\end{figure}

In figure \ref{kin} kinematic parameters are presented. These
parameters are derived from the spectra, using a $\chi^2$ minimization
technique in order to find the parameters of a truncated Gauss-Hermite
series that gives the best approximation to the LOSVD.  The spectra
were spatially rebinned to satisfy a minimum S/N. Only data points
from the side of the galaxy with mean streaming away from the observer
are shown. A best fitting template mix (i.e. the one that gives the
smallest $\chi^2$ in the fitting) was determined for the central data
point and this mix was then used throughout the whole galaxy.  The
spectra are analysed with a mix of a G5V dwarf and a K4III giant.

The kinematic parameters are determined following the method of
\citet{vdmf} and implemented as in \citet{rijdi}, first fitting a
Gaussian LOSVD to determine $\langle v\rangle$ and $\sigma$, and then
keeping these lowest order parameters fixed to fit the other
coefficients.

If $\langle v\rangle$, $\sigma$, $h_3$ and $h_4$ are the lowest order
parameters for the LOSVD, the true moments are approximated more
accurately by
\begin{equation}
\langle v_p\rangle = \langle v\rangle + \sqrt{3}\sigma h_3
\end{equation}
for the first order moment
and 
\begin{equation}
\sigma_p = \sigma(1+\sqrt{6} h_4)
\end{equation}
for the second order moment,
$\langle v_p\rangle$ and $\sigma_p$ are also called the 'corrected' values.

These data are consistent with the kinematic parameters presented by \citet{pel} and \citet{kop}.


\begin{table}
\begin{tabular}{l|l}
\hline
HD 21019 & G2V\\
HD 115617&G5V\\
HD 102070 & G8III\\
HD 117818&K0III\\
HD 114038&K1III\\
HD 44951 & K3III\\
HD 29065&K4III\\
HD 99167 & K5III\\
HD 129902&M1III\\
HD 93655&M2III\\
\hline
\end{tabular}
\caption{Observed stellar templates.}
\label{temps}
\end{table}

\subsection{Other observations}
NGC~3258 is a weak but extended radio source with a total flux of 18.3
mJy (VLA observations at 4.9 GHZ (6 cm) within $45''$) \citep{sad},
where the core ($< 5''$) flux is only $7\%$ of the total flux.  A hot
gaseous corona around NGC~3258 emits about $3\times 10^{41} \rm{ergs}
s^{-1}$ \citep{fab}. There are no total mass estimates based on X-ray
observations available.

\citet{breg} studied this galaxy in the far-infrared, and derived a
dust temperature of 24.5 K and a dust mass of $6\times 10^6 M_\odot$
from IRAS observations at 60 and $100\mu\rm{m}$. This is not unlike
what is found in spiral galaxies.  The mid-infrared properties of
NGC~3258, based upon ISOCAM observations at 6.75, 9.63 and
$15\mu\rm{m}$, indicate the presence of a hot $T\simeq 260$K) dust
component \citep{fer}. A mass of $48 M_\odot$ of hot dust within a
radius of 5.5 kpc was derived. This is at the lower end of the
typical range of 10-400 $M_\odot$ reported for early-type galaxies.


Furthermore, the galaxy shows [OII] emission \citep{ber} and weak HI
emission \citep{jen}. 

\section[]{Dynamical modelling strategy}

There is a large variety of ways to make a dynamical model and there
also different strategies to use these models. For a recent review,
see \citet{job}.


The data used for the modelling are the projected surface brightness
and a set of spectra of the galaxy, taken at several positions along
the major axis. The projected surface brightnes is used to obtain a
graviational potential, this is described in more detail in \ref{pot}.



An important issue for the modelling strategy applied in this paper,
is the form of the data that are used to determine the DF.  In this
case, the fit is done directly to the spectra (after necessary data
reduction steps). This means that the observed quantities that have to
be reproduced by the model are pairs (wavelength, flux).  The use of
spectra as input for dynamical information is in contrast with more
frequently used modelling techniques where a number of kinematic
parameters are first derived from the spectra. The models are then
fitted to these parameters that are moments of the distribution
function. The modelling procedure used in this paper is described in
detail in DD98.  The input spectra are reproduced by a dynamical
model that uses spectra of different template stars. An important
characteristic of these models is that they provide a DF for each
population that is used in the model.

This is a function which gives the probability to find a star on an
orbit characterized by the integrals of motion $E$ (the energy), $L$
(the total angular momentum) and $L_z$ (the component of the angular
momentum along the rotation axis of the system). This function holds
all information on the dynamical state of the galaxy.

The resulting DF for each type of star is a sum of basic components
and is constructed with a quadratic programming algorithm
\citep{dj}. There are two main requirements for the DF: (1) the
difference between the observed quantities and the quantities
calculated from the DF should be minimized (least squares
minimization) and (2) the DF should be a positive function.

Modelling the spectra instead of kinematic parameters causes a
considerable increase in the number of data points and hence an
increase of the required computing time, but there are a number of
advantages over using kinematic parameters.

\begin{itemize}
\item More traditional modelling involves a two step process
(e.g. \citet{db}): first determining kinematic parameters out of
observed spectra and then calculating a model. Here this is turned
into a one step process.  Omitting the step of determining kinematic
parameters and using the galactic spectra as input has the consequence
that the kinematic information used as input for the modelling is
independent of any parameterization. Also in the more traditional two
step modelling process, there are some non-parametric methods to
derive kinemtical profiles, see \citet{db2}.

Of course, the dynamical profiles that are a result from the modelling
depend on the assumptions concerning the gravitational
potential and the dark matter halo, but this is the case for every
dynamical modelling technique. In this paper, the gravitational
potential is assumed to be spherical. This assumption is not a
conceptual issue but is justified by the shape of the NGC~3258.

\item Furthermore, since the basis distribution functions that are
used in the fit are chosen freely by the algorithm from a large
library, also the modelling procedure itself is non-parametric. The
smoothness of the model is guaranteed by the use of smooth basis
functions. Hence, considering also the previous remark, this method
can be regarded as a non-parametric modelling method. Moreover, the
model has considerably more freedom in determining LOSVDs than can be
achieved with a limited number of parameters as is the case when
kinematic parameters are determined in a separate process.

\item The modelling process based on quadratic programming, uses a
$\chi^2$ (see equation (10) in DD98) that has the statistical
meaning of a goodness of fit indicator if the noise on the spectra is
Poisson noise and the data points are weighted accordingly.

\item The Hessian matrix (see equation (13) in DD98) involved in
the quadratic programming can be used to calculate error bars on the
distribution functions that have a statistical meaning and give an
idea of the uncertainty on the model and on its moments.  Tests with
synthetic spectra (DD98) show that the error bars are a reliable
measure for the spread between input model and fit, as caused by the
adopted Poisson noise.  Moreover, if the component library is
sufficiently complete the systematic errors are negligible compared to
the random errors.

\item If the dynamical modelling uses different template stars, it can
result in a population synthesis. 

\item The composition of the stellar mix is allowed to vary smoothly
as a function of radius and is therefore directly connected to the
internal dynamics. This minimizes stellar template mismatch, and
allows in meantime to obtain a physical stellar population
decomposition.

\item Information at different wavelengths can be combined in an
elegant way. This is an advantage of using kinematic information on
the level of spectra. It is not clear how to interpret profiles of the
same kinematic parameters taken from different parts of the spectrum
that give different results \citep{vdm}.

\end{itemize}

\subsection{Templates}
Five template stars that have discernible spectral properties were
used: a G2V dwarf, a G5V dwarf, a K1III giant, a K4III giant and a
M1III giant (see table \ref{temps} for their identification).

The dynamical models were created in an iterative process.  First,
models were calculated with a single template star, using a library of
about 100 dynamical components.  Afterward, the components that are
selected for the single template models are combined into component
libraries that contain a sample of dynamical components for different
stellar templates. With these mixed libraries, new models were
calculated.

\section{Dynamical components}
Using a spherical potential in combination with rotating components
for the DF, it is possible to obtain a flattened luminous density
distribution, as is required to be able to fit the two dimensional
surface brightness of this E1 galaxy well. We consider three integral
models: $F(E,L,L_z)$ for which $\sigma_r$, $\sigma_\phi$ and
$\sigma_\theta$ are different. These models are linear combinations of
three types of components.

\subsection{Continuum terms}
The continuum of the weighted sum of template spectra does not
necessarily fit the continuum of the galaxy spectra. Therefore, low
order polynomials are included in the component library to compensate
for the small differences between both continua. Such a continuum
component has the form
\begin{equation}
g(\ln\lambda,r_p) = \left({\ln \lambda -\ln \lambda_1}\over{\ln
\lambda_2-\ln \lambda_1}\right)^\alpha
\left({j_R(r_p)-j_R(r_c)}\over{j_R(0)}\right)^\beta,
\end{equation}
with $\alpha$ and $\beta$ real numbers, $r_c$ the cutoff radius (the
component is set to zero for larger radii), $j_R(0)$ the central value
of the surface brightness, $j_R(r_c)$ the surface brightness at the
cutoff radius, $j_R(r_p)$ the surface brightness at the position of
the line of sight, $\lambda$ the wavelength at which the spectrum is
evaluated, $\lambda_1$ and $\lambda_2$ bound the spectral region that
is used in the fit. The value of $\alpha$ should be small so that these
components only fit the continuum and not the absorption lines.

\subsection{Fricke components}
These components have the following augmented density
\begin{equation}
\tilde{\rho}(\psi,r) = (\psi-E_0)^\alpha \left(r\over r_c \right)^{2\beta}, \quad \psi\ge E_0, \; 0 \; {\rm elsewhere}
\label{aug1}
\end{equation}
with $\alpha$ and
$\beta$ positive integers, $r_c$ a scale length and $E_0$ the energy
level at which the component is truncated.  
All of these components are either isotropic ($\beta=0$) or
tangentially anisotropic (component parameter $\beta>0$). However, the
coefficients for these components can be negative, hence also radial
anisotropic models can be constructed. These are spherical
two integral $(E,L)$ components. They have equal velocity dispersions
in the tangential directions, $\sigma_\phi=\sigma_\theta$, but a
different radial dispersion in the case of non-isotropic components.
A more detailed discussion of this family of components can be found
in DD98.

\subsection{Rotating components}

These components add rotation to the model. Due to the ordered motion
of the stars around the component's symmetry-axis, the tangential
dispersion is usually lower than the radial dispersion. These rotating
components are axisymmetric two integral $(E,L_{z'})$ components (with
$z'$ the symmetry axis of the component), for which $\sigma_R=\sigma_{z'}$
and the dispersion in the rotation direction $\sigma_\phi$ is
different.

In this section, a family of components is introduced that should
enable one to reproduce both the rotation of the bulk of the galaxy
and the kinematics of a small subsystem, rotating around an axis other
than the main galaxy's rotation axis.  The distribution function of
the components of this family is given by
\begin{eqnarray}
F(E,L_{z'}) &=& \frac{1}{2^{\beta/2}} (E-E_0)^{\alpha}
\left( i_{\rm rot} L_{z'}- L_0 \right)^{\beta} \nonumber \\
&& \hspace{.5em} \mbox{for }  E >E_0 \nonumber \\
&& \hspace{.5em} \mbox{and }   i_{\rm rot} = 1,\,L_{z'}>0 \mbox{ and } L_{z'} > L_0 \nonumber \\
&& \hspace{.5em} \mbox{or } i_{\rm rot} = -1,\,L_{z'}<0 \mbox{ and }  
|L_{z'}|> L_0, \label{DF2}
\end{eqnarray}
with the energy per unit mass $E = \psi(r) - v_M^2/2 -
L_{z'}^2/(2\varpi^2)$ and the $z'$-component of the angular
momentum $L_{z'} = \varpi v_\phi$. The radial distance to the centre
is measured by $r$, the distance to the centre measured in the
equatorial plane is denoted by $\varpi$. The velocity component in the
meridional plane is $v_M = \sqrt{v_\varpi^2 + v_{z'}^2}$, the velocity
component perpendicular to the meridional plane is given by
$v_\phi$. Here, $(x',y',z')$ is a Cartesian reference frame, attached
to the component, with the $z'$-axis the rotation axis of the
component. The overall gravitational potential $\psi(r)$ is taken to
be spherical symmetric although the components themselves have an
axisymmetric mass distribution. This is acceptable since a small
kinematically distinct core probably does not significantly influence
the global potential. Also, a slowly rotating elliptical galaxy can
still be approximately round, justifying the use of a spherical
potential. An axisymmetric potential generally allows $E$ and $L_{z}$
as integrals of motion, with $L_z$ the component of the angular
momentum parallel to the potentials symmetry axis. Therefore, these
components can also be used in an axisymmetric potential if their
rotation axis $z'$ coincides with the symmetry axis $z$ of the
potential.

All the stars have the same sense of rotation, i.e. if $i_{\rm rot}=1$
the stars rotate counter-clockwise, if $i_{\rm rot}=-1$ they rotate
clockwise. The parameter $\alpha$ determines the distribution of the
stars over orbits with different energies whereas $\beta$ influences
the way orbits with different angular momenta are populated. The
energy cutoff $E > E_0$ will prevent stars from reaching beyond a
certain maximum radius. The angular momentum cutoff $|L_{z'}|> L_0$ on
the other hand will impose a minimum radius that stars can reach.

\subsubsection{The component's frame of reference}

These components are to be employed in the modelling of slowly
rotating, approximately spherical galaxies using a severally
symmetric potential. Throughout this work, all dynamical quantities
are expressed in spherical coordinates. These particular components
however are axisymmetric and their characteristics will be most
readily calculated in cylindrical coordinates. Therefore, we seek a
transformation linking the spherical coordinates in which the
properties of the galaxy as a whole are expressed to a cylindrical
coordinate system, attached to each individual component of this
family.

First, we introduce a cartesian reference frame ($x,y,z$) with its
origin at the galaxy's centre. The components of a vector $\vec{a}$ in
spherical coordinates $(r,\theta,\varphi)$ can be linked to its
components in cartesian coordinates via the transformation
\begin{equation}
\left( \begin{array}{c}
	a_r \\
	a_\theta \\
	a_\varphi
       \end{array} \right) 
 = {\bf A} \left( \begin{array}{c}
	a_x \\
	a_y \\
	a_z
       \end{array} \right),
\end{equation}
with
\begin{equation}
 {\bf A}= 
\left( \begin{array}{ccc}
        \sin \theta \cos \varphi & \sin \theta \sin \varphi & \cos \theta \\
	\cos \theta \cos \varphi & \cos \theta \sin \varphi & - \sin \theta \\
	-\sin \varphi & \cos \varphi & 0 
       \end{array} \right)
\end{equation}
Suppose the direction of the rotation axis of the system we are
describing, whether it be a small subsystem or the galaxy's main body,
is given by the angles $(\theta_0,\varphi_0)$ with respect to the
cartesian reference frame. The cartesian coordinate system
$(x',y',z')$, fixed to the component, can be obtained by first
rotating the $(x,y,z)$-frame over an angle $\varphi_0$ around the
$z$-axis, yielding the auxiliary $(x'',y'',z'')$-frame, followed by a
rotation of the $(x'',y'',z'')$-frame over an angle $\theta_0$ around
the $y''$-axis :
\begin{eqnarray}
\left( \begin{array}{c}
	a_x \\
	a_y \\
	a_z
       \end{array} \right) =
 {\bf B} \left( \begin{array}{c}
	a_{x'} \\
	a_{y'} \\
	a_{z'}
       \end{array} \right),
\end{eqnarray}
with
\begin{eqnarray}
 {\bf B}=
\left( \begin{array}{ccc}
	\cos \theta_0 \cos \varphi_0 & -\sin \varphi_0 & \sin \theta_0 \cos \varphi_0 \\
	\cos \theta_0 \sin \varphi_0 & \cos \varphi_0 & \sin \theta_0 \sin \varphi_0 \\
	- \sin \theta_0	& 0	& \cos \theta_0
       \end{array} \right) .
\end{eqnarray}

To facilitate the calculations as much as possible, we will henceforth
use cylindrical coordinates $(\varpi,\phi,z')$. These are connected to
the $(x',y',z')$-frame by the transformation
\begin{eqnarray}
\left( \begin{array}{c}
	a_{x'} \\
	a_{y'} \\
	a_{z'}
       \end{array} \right) &=&
\left( \begin{array}{ccc}
	\cos \phi & - \sin \phi & 0 \\
	\sin \phi & \cos \phi & 0 \\
	0	  &	0     & 1
\end{array} \right) 
\left( \begin{array}{c}
	a_\varpi \\
	a_\phi \\
	a_{z'}
       \end{array} \right)\\
 &=& {\bf C} \left( \begin{array}{c}
	a_\varpi \\
	a_\phi \\
	a_{z'}
       \end{array} \right).
\end{eqnarray}
One can now immediately make the connection between the spherical
coordinates $(r,\theta,\varphi)$ in the galaxy's reference frame and
the cylindrical coordinates $(\varpi,\phi,z')$ in the component's frame of
reference using the matrix ${\bf D} = {\bf ABC}$ :
\begin{equation}
\left( \begin{array}{c}
	a_r \\
	a_\theta \\
	a_\varphi
       \end{array} \right) = {\bf D} 
\left( \begin{array}{c}
	a_\varpi \\
	a_\phi \\
	a_{z'}
       \end{array} \right).
\end{equation}

\subsubsection{The velocity moments}

The axisymmetric velocity moments for a counter-clockwise rotating
model ($L_{z'} \ge 0$, $i_{\rm rot} =1$) are defined as
\begin{eqnarray}
&&\tilde{\mu}^+_{l,m,n}(\psi,\varpi) = \nonumber \\&& \int_{-\infty}^{\infty}
\int_{-\infty}^{\infty} \int_{-\infty}^{\infty} F(E,L_{z'})\, v^l_\varpi
v^m_\phi v^n_{z'}\,dv_\varpi\,dv_\phi\,d_{z'}
\end{eqnarray}
with $E=\psi(\sqrt{\varpi^2+z'^2}) - \frac{v^2}{2}$ and $L_{z'} =
\varpi v_\phi$. The moments of a clockwise rotating model 
($L_{z'} \le 0$, $i_{\rm rot}=-1$) with the same parameters $\alpha$
and $\beta$ are readily obtained since
\begin{equation}
\tilde{\mu}^-_{l,m,n}(\psi,\varpi) = i_{\rm rot}^m \times
\tilde{\mu}^+_{l,m,n}(\psi,\varpi).
\end{equation}
The distribution function is an even function of $v_\varpi$ and
$v_{z'}$, therefore only the moments with even $l$ and $n$ values will
be non-zero. One can define anisotropic moments
\begin{equation}
\tilde{\mu}^+_{2n,m}(\psi,\varpi) = 2 \pi \int_0^{+ \infty} dv_M \int_{-
\infty}^{+ \infty} dv_\phi \, F(E,L_z) v_\phi^m v_M^{2n+1},
\end{equation}
that are connected to the true moments by the relation
\begin{equation}
\tilde{\mu}^+_{2l,m,2n}(\psi,\varpi) = \frac{1}{\pi} B(l + \frac{1}{2},
n + \frac{1}{2}) \tilde{\mu}^+_{2(l+n),m}(\psi,\varpi). \label{aximomani}
\end{equation}

Performing some calculations the expression for the
anisotropic velocity moments becomes:
\begin{eqnarray}
\tilde{\mu}^+_{2n,m}(\psi,\varpi) &=& 2^{\alpha+2n+(m+5)/2} \pi
 \Gamma(\alpha+1) \Gamma(n+1) \\ &&\hspace{-7em}\times\varpi^\beta (\psi-E_0)^{(\alpha+n+1)/2}
  \nonumber \\ &&  \hspace{-7em}\times \sum_{i=0}^{m} {m \choose i}
 \frac{\Gamma(\beta+i+1)}{\Gamma(\alpha+\beta+n+i+3)} \left(
 \frac{L_0}{\sqrt{2}\varpi} \right)^{m-i} \nonumber \\
&&  \hspace{-7em}\times \left( \sqrt{\psi-E_0}-
 \frac{L_0}{\sqrt{2}\varpi} \right)^{\alpha+\beta+n+i+2} 
 \nonumber \\&& \hspace{-7em}\times \,_2F_1 \left(
 -\alpha-n-1,\alpha+n+2;\alpha+\beta+n+i+3;\right.\nonumber\\
&& \hspace{-6em}\left.\frac{1}{2}
 \frac{\sqrt{2\varpi^2(\psi-E_0)}- L_0}{\sqrt{2\varpi^2(\psi-E_0)}}
 \right).
\end{eqnarray}
The true velocity moments then immediately follow :
\begin{eqnarray}
\tilde{\mu}_{l,m,n}(\psi,\varpi)
 &=& i_{\rm rot}^m
2^{\alpha+l+n+(m+5)/2} \nonumber\\
&&\hspace{-7em} \times\Gamma(\alpha+1) \Gamma(\frac{l+1}{2})
\Gamma(\frac{n+1}{2}) \varpi^\beta
  \nonumber \\ &&
\hspace{-7em} \times (\sqrt{\psi-E_0})^{\alpha+(l+n)/2+1} \nonumber\\
&& \hspace{-7em} \times\sum_{i=0}^{m} {m
 \choose i}
 \frac{\Gamma(\beta+i+1)}{\Gamma(\alpha+\beta+\frac{l+n}{2}+i+3)}
 \left( \frac{L_0}{\sqrt{2}\varpi} \right)^{m-i} \nonumber \\&&
 \hspace{-7em} \times\left( \sqrt{\psi-E_0}- \frac{L_0}{\sqrt{2}\varpi}
 \right)^{\alpha+\beta+(l+n)/2+i+2} \nonumber \\ &&
 \hspace{-7em} \times \,_2F_1 \left(
 -\alpha-\frac{l+n}{2}-1,\alpha+\frac{l+n}{2}+2;\right.\nonumber\\
&&\hspace{-6em}\left.
\alpha+\beta+\frac{l+n}{2}+i+3;
\frac{\sqrt{2\varpi^2(\psi-E_0)}-
 L_0}{2\sqrt{2\varpi^2(\psi-E_0)}} \right).
\label{momfam2}
\end{eqnarray}
 
\subsubsection{Some commonly used moments}
\paragraph{Spatial density}

The spatial density is given by
\begin{eqnarray}
\lefteqn{ \rho(\varpi,z') = \mu_{0,0,0}(\varpi,z') = } \nonumber \\
&&  2^{\alpha+5/2} \pi \varpi^\beta \frac{\Gamma(\alpha+1)
 \Gamma(\beta+1)}{\Gamma(\alpha+\beta+3)}\nonumber\\
&&\hspace{-.5em}\times
\left(\sqrt{\psi(r)-E_0}\right)^{\alpha+1} \left(\sqrt{\psi(r)-E_0}-
\frac{L_0}{\sqrt{2}\varpi} \right)^{\alpha+\beta+2}   \nonumber \\
&& \hspace{-.5em}\times\,_2F_1 \left(-\alpha-1,\alpha+2;\alpha+\beta+3;\right.\nonumber\\
&& \left.
\frac{\sqrt{2\varpi^2(\psi-E_0)}-L_0}{2\sqrt{2\varpi^2(\psi-E_0)}} \right). 
\label{densfam2}
\end{eqnarray}

\begin{figure}
\includegraphics[bb=290 35 540 725,angle=-90,scale=.35,clip=true]{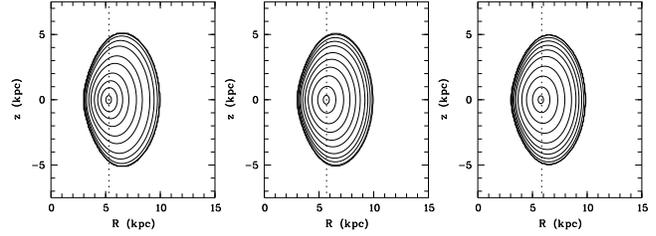}
\caption{A plot of the spatial density in the meridional plane of 
models with $r_{\rm min}=3$~kpc, $r_{\rm max}=10$~kpc, $\alpha=4$ and
$\beta=4$ (left), 8 (middle) and 12 (right). The boundary of this
region is given by equation \ref{rand}, hence all models fill the same volume. The
higher the value of $\beta$, the more the high-angular momentum orbits
are populated. This causes the position of the peak density (marked by
the dashed line) to slowly shift outwards. }
\label{spadens}
\end{figure}

The density is only non-zero inside the volume defined by $\psi(r)-E_0
> L_0^2/(2\varpi^2)$. Evidently, the boundary of this volume,
where the density becomes exactly zero, is determined by
\begin{equation}
\psi(r) = E_0 + \frac{L_0^2}{2 \varpi^2}. \label{rand}
\end{equation}
For non-zero $L_0$ and $E_0$, this volume has approximately the shape
of a torus. If $L_0=0$, the density is non-zero inside the spherical
region defined by $r \le \psi^{-1}(E_0)$. All models that share the
same values for $E_0$ and $L_0$ fill the same volume, independent of
$\alpha$ or $\beta$ but the density distribution inside this volume
does depend on $\alpha$ and $\beta$, see figure \ref{spadens}  

\begin{figure}
\includegraphics[bb=50 50 320 780, angle=0,scale=.5,clip=true]{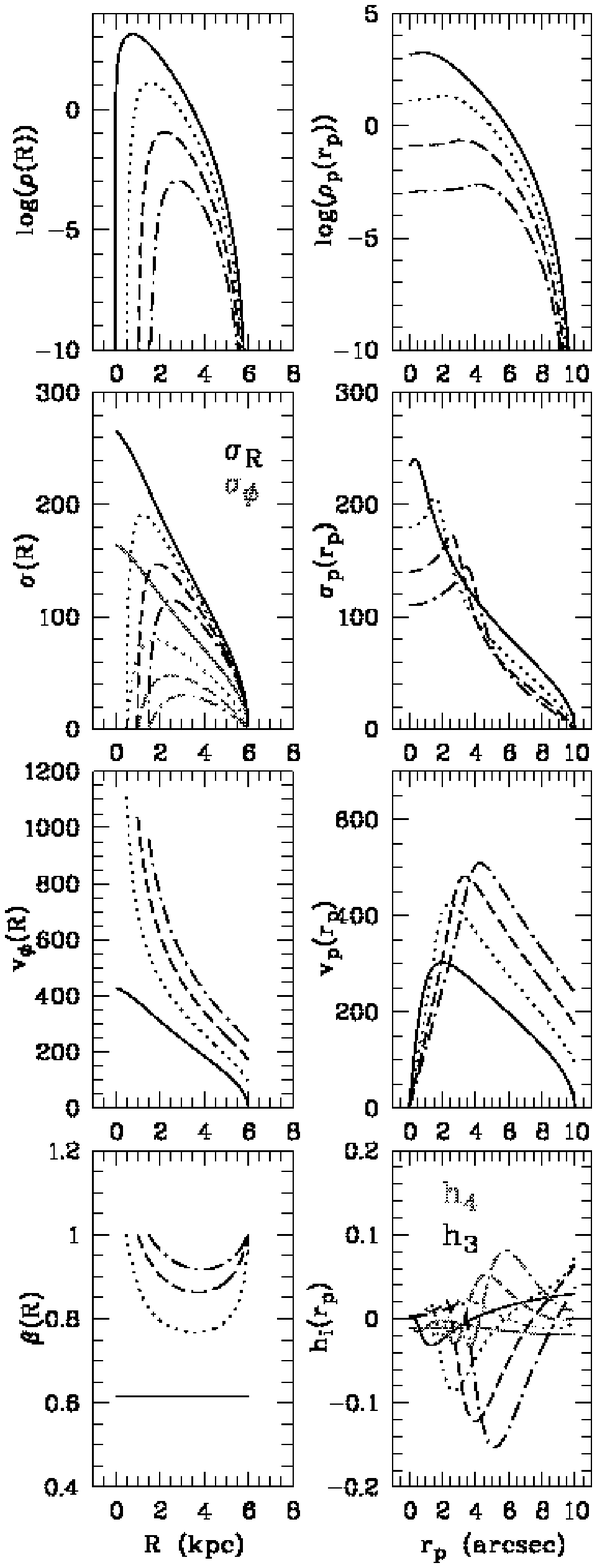}
\caption{The influence of the angular momentum cut-off on 
the kinematics of models with $\alpha=6$, $\beta=2$, $r_{\rm
max}=6$~kpc is demonstrated. The values of the inner radius are
$r_{\rm min}=0$~kpc (solid line), 0.5~kpc (dotted line), 1.0~kpc
(dashed line) and 1.5~kpc (dashed-dotted line). In the left column, a
number of intrinsic kinematic quantities are plotted. The radius $R$
is measured in the equatorial plane, $z'=0$. From top to bottom : the
logarithm of the spatial luminosity density $\rho(R)$, the radial
(black) and the tangential velocity dispersion (gray), the rotation
velocity $v_\phi(R)$ and Binney's anisotropy parameter $\beta(R)$. In
the right column, the projected kinematics are presented. The radius
$r_p$ lies along the long axis of the models. From top to bottom : the
logarithm of the projected luminosity density $\rho_p(r_p)$, the
projected velocity dispersion $\sigma_p(r_p)$, the projected streaming
velocity $v_p(r_p)$ and the Gauss-Hermite coefficients $h_3$ and
$h_4$ (gray).}
\label{cutfam2} 
\end{figure}

\begin{figure}

\includegraphics[bb=50 50 320 780,scale=.5,clip=true]{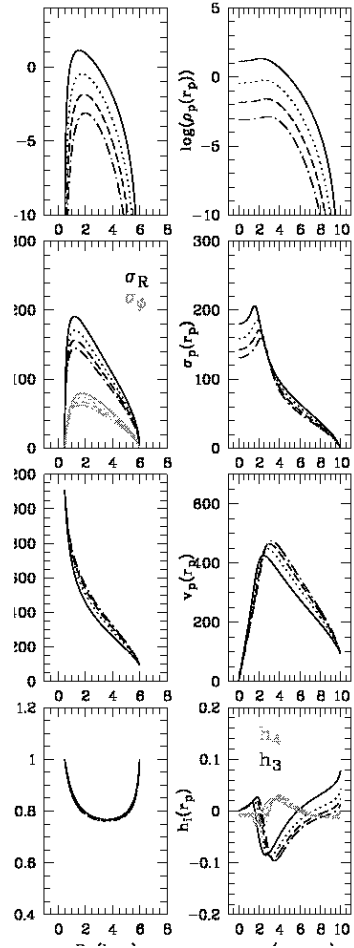}
\caption{The influence of the parameter $\beta$ on 
the kinematics of models with $\alpha=6$, $r_{\rm min}=0.5$~kpc and
$r_{\rm max}=6$~kpc is demonstrated. The values of $\beta$ are 2 (full
line), 4 (dotted line), 6 (dashed line) and 8 (dashed-dotted line). In
the left column, a number of intrinsic kinematic quantities are
plotted. The radius $R$ is measured in the equatorial plane,
$z'=0$. From top to bottom : the logarithm of the spatial density
$\rho(R)$, the radial (black) and the tangential velocity dispersion
(gray), the rotation velocity $v_\phi(R)$ and Binney's anisotropy
parameter $\beta(R)$. In the right column, the projected kinematics
are presented. The radius $r_p$ lies along the long axis of the
models. From top to bottom : the logarithm of the projected density
$\rho_p(r_p)$, the projected velocity dispersion $\sigma_p(r_p)$, the
projected streaming velocity $v_p(r_p)$ and the Gauss-Hermite
coefficients $h_3$ (black) and $h_4$ (gray).}
\label{beta}
\end{figure}
\paragraph{Mean velocity}

For the mean rotation velocity, one finds
\begin{eqnarray}
v_\phi(\varpi,z') &=& 
       \frac{\mu_{0,1,0}(\varpi,z')}{\rho(\varpi,z')} \nonumber
       \\ && \hspace{-7em}= \frac{L_0}{\varpi} + \frac{\beta+1}{\alpha+\beta+3}
       \left(\sqrt{\psi(r)-E_0}-\frac{L_0}{\sqrt{2}\varpi}
       \right)  \nonumber \\ &&\hspace{-7em}\times\frac{\,_2F_1 \left(
       -\alpha-1,\alpha+2;\alpha+\beta+4;\frac{\sqrt{2\varpi^2(\psi-E_0)}-
       L_0}{2\sqrt{2\varpi^2(\psi-E_0)}} \right)}{\,_2F_1 \left(
       -\alpha-1,\alpha+2;\alpha+\beta+3;\frac{\sqrt{2\varpi^2(\psi-E_0)}-
       L_0}{2\sqrt{2\varpi^2(\psi-E_0)}} \right)}.
\end{eqnarray}
At the boundary of the component, the mean rotation velocity takes the
value $v_\phi = L_0/\varpi = \sqrt{2(\psi(r)-E_0)}$. The
parameter $\beta$ has little impact on the rotation velocity profile,
as shown in figure \ref{cutfam2} and figure \ref{beta}. The extreme values of
the rotation velocity are determined solely by $L_0$ and $E_0$, making
the rotation velocity a very sensitive function of these parameters.

\paragraph{Velocity dispersion}

Since $l$ and $n$ are completely equivalent in equation
(\ref{momfam2}), it is obvious that the velocity dispersions
$\sigma^2_\varpi(\varpi,z')$ and $\sigma^2_{z'}(\varpi,z')$ are always
equal. This is the case for any model with a distribution function
that depends only on $E$ and $L_{z'}$. One finds that
\begin{eqnarray}
\sigma^2_\varpi(\varpi,z') &=&  \sigma^2_{z'}(\varpi,z')  \nonumber\\&&\hspace{-6em}
=\frac{2}{\alpha+\beta+3} \sqrt{\psi(r)-E_0}
\left(\sqrt{\psi(r)-E_0}-\frac{L_0}{\sqrt{2}\varpi} \right) 
\nonumber \\
&&\hspace{-6em}\times \frac{\,_2F_1 \left(
       -\alpha-2,\alpha+3;\alpha+\beta+4;\frac{\sqrt{2\varpi^2(\psi-E_0)}-
       L_0}{2\sqrt{2\varpi^2(\psi-E_0)}} \right)}{\,_2F_1 \left(
       -\alpha-1,\alpha+2;\alpha+\beta+3;\frac{\sqrt{2\varpi^2(\psi-E_0)}-
       L_0}{2\sqrt{2\varpi^2(\psi-E_0)}} \right)} \\
\sigma^2_\phi(\varpi,z') &=& 2 \sum_{i=0}^2 
\frac{(\beta+1)_i}{(\alpha+\beta+3)_i} 
\left( \frac{L_0}{\sqrt{2}\varpi} \right)^{2-i}
\nonumber\\&&\hspace{-6em}\times\left(\sqrt{\psi(r)-E_0}-\frac{L_0}
{\sqrt{2}\varpi} \right)^i  \nonumber \\
&&\hspace{-6em} \times\frac{\,_2F_1 \left(
       -\alpha-1,\alpha+2;\alpha+\beta+i+3;\frac{\sqrt{2\varpi^2(\psi-E_0)}-
       L_0}{2\sqrt{2\varpi^2(\psi-E_0)}} \right)}{\,_2F_1 \left(
       -\alpha-1,\alpha+2;\alpha+\beta+3;\frac{\sqrt{2\varpi^2(\psi-E_0)}-
       L_0}{2\sqrt{2\varpi^2(\psi-E_0)}} \right)}\nonumber\\
&& \hspace{-6em}- v^2_\phi(\varpi,z').
\end{eqnarray}

Figures \ref{cutfam2} and \ref{beta} also show that the anisotropy
parameter $\beta_{\phi}$, with
\begin{equation}
\beta_phi(\varpi,z')=1-\frac{\sigma_\phi(\varpi,z')}{\sigma_\varpi(\varpi,z')}
\end{equation}
is almost independent of the components parameter $\beta$
but very sensitive to the value of the angular momentum cutoff
$L_0$. 

\subsubsection{The spatial line-of-sight velocity distribution}

For the calculation of the spatial LOSVD, the DF will be directly
integrated over the two velocity components in the plane of the
sky. The integration over the line of sight will have to be performed
numerically since it depends on the shape of the potential, which is
generally only known numerically. We decompose the velocity vector in
a component parallel to the line of sight, $v_p$, and two components
in the plane of the sky, $v_1$ and $v_2$. The polar unit vector
$\vec{e}_\phi$ can be written with respect to the vectors
($\vec{e}_z,\vec{e}_1,\vec{e}_2$) with $\vec{e}_g$ the unit vector
along the line of sight and $\vec{e}_1$ and $\vec{e}_2$ unit vectors
in the plane of the sky :
\begin{equation}
\vec{e}_\phi = b_1 \vec{e}_g + b_2 \vec{e}_1 + b_3 \vec{e}_2,
\end{equation}
with $b_1^2 + b_2^2 + b_3^2=1$. We now rewrite the distribution
function as a function of $v_p$, $v_1$ and $v_2$ :
\begin{eqnarray}
\lefteqn{F(\varpi,z, v_\varpi, v_\phi, v_z)\,dv_\varpi\,dv_\phi\,dv_z 
= F(\varpi,z, v_p,v_1,v_2)\,dv_p\,dv_1\,dv_2 } \nonumber \\
&=& 
\frac{\varpi^n}{2^{n/2}} \left(\psi-E_0-\frac{1}{2}(v_p^2+v_1^2+v_2^2)\right)^\alpha 
 \nonumber \\ &\times&
\left(i_{\rm rot} (b_1 v_p + b_2 v_1 + b_3 v_2) - \frac{L_0}{\varpi}\right)^n
\,dv_p\,dv_1\,dv_2.
\label{draai3}
\end{eqnarray}
Here and in the following, the power on the angular momentum will be
assumed to be an integer, denoted by $n$. LOSVDs of components with
non-integer values for this parameter will have to be calculated
numerically. The velocity components in the plane of the sky, $v_1$
and $v_2$, can be expressed in polar coordinates $(v,\vartheta)$
\begin{eqnarray}
v_1 &=& v \cos \vartheta \hspace{3em} v \ge 0,\,\, 
                                      \vartheta \in [0,2\pi] \nonumber \\
v_2 &=& v \sin \vartheta.
\end{eqnarray}
Taking this transformation into account and introducing the notation
\begin{eqnarray}
b_2 &=& b \cos \chi \hspace{3em} b \ge 0,\,\, 
                                 \chi \in [0,2\pi] \nonumber \\
b_3 &=& b \sin \chi,
\end{eqnarray}
following expression can be obtained
\begin{eqnarray}
F(\varpi,z,v,\vartheta,v_p) \,dv_p v\,dv\,d\vartheta  &=& \frac{\varpi^n}{2^{n/2}}
\left(\psi-E_0-\frac{1}{2}(v_p^2+v^2)\right)^\alpha  \nonumber \\
&& \hspace{-12em} \times\left(i_{\rm rot} b_1 v_p + b v \cos \vartheta - \frac{L_0}{\varpi} \right)^n \,dv_p v\,dv\,d\vartheta,
\end{eqnarray}
in which $\chi$ has been dropped, since the sign and the zero-point of
$\vartheta$ are immaterial. The cosines $b$ and $b_1$ obey the
relation $b^2 + b_1^2 = 1$, with $b$ always positive. The distribution
function has to be integrated over that part of the rectangle
\begin{eqnarray}
0 \le &v& \le \sqrt{2(\psi-E_0)-v_p^2}, \nonumber \\
0 \le &\vartheta& \le 2 \pi \label{rect}
\end{eqnarray}
where 
\begin{equation}
i_{\rm rot} b_1 v_p + b v \cos \vartheta - \frac{L_0}{\varpi} \ge 0. 
\label{constra}
\end{equation}
To relieve the notation throughout the remaining calculations, the
following parameters will be used
\begin{eqnarray}
{\cal A} &=& \sqrt{2(\psi-E_0)-v_p^2}, \nonumber \\
{\cal P} &=& i_{\rm rot} b_1 v_p - \frac{L_0}{\varpi}, \nonumber \\
\cos \vartheta_0 &=& - \frac{{\cal P}}{{\cal A} b}.
\end{eqnarray}

The calculations will proceed along different paths depending on
whether ${\cal P} \ge 0$ or ${\cal P} \le 0$. The area in the
rectangle (\ref{rect}) over which the distribution function has to be
integrated is further limited by the positivity constraint on the
angular momentum, (\ref{constra}). This translates in the extra
constraints
\begin{figure}
\includegraphics[bb=50 55 545 730,angle=-90,scale=.4,clip=true]{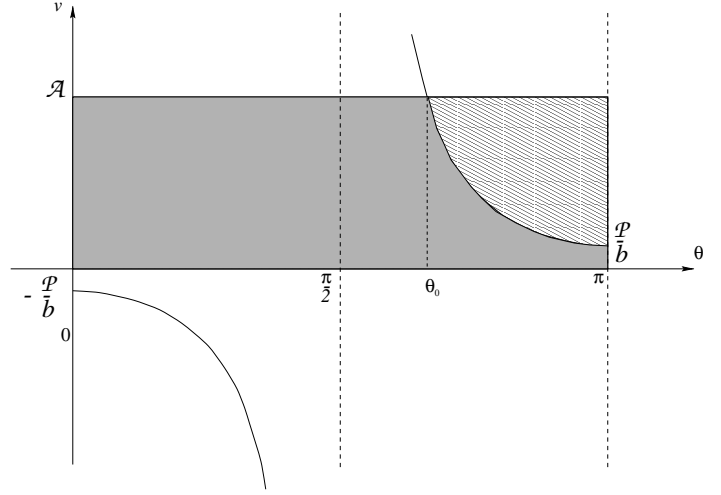}
\caption{The grey shaded area in the $(v,\vartheta)$-plane marks 
the area over which the distribution function has to be integrated to
calculate the spatial LOSVD if ${\cal P}
\ge 0$. The hatched rectangle is defined by $0 \le v
\le \cal A$ and $0 \le \vartheta \le \pi$. The integration area is
bounded at the right by the curve $v \le \frac{-\cal P}{b \cos
\vartheta}$. \label{fam2int1}}
\end{figure}
\begin{eqnarray}
v &\ge& - \frac{{\cal P}}{b \cos \vartheta},\,\,\,\cos \vartheta \ge
0, \nonumber \\ v &\le& - \frac{{\cal P}}{b \cos \vartheta},\,\,\,\cos
\vartheta \le 0. \label{curv}
\end{eqnarray}
The spatial LOSVD can then be calculated as
\begin{eqnarray}
\phi(\varpi,z',v_p) &=& \frac{\varpi^n}{2^{\alpha + n/2}} \int_0^{{\cal A}^2} 
({\cal A}^2 - v^2)^\alpha dv^2 \nonumber\\
&&\times\int_0^\pi \left( {\cal P} + b v \cos
\vartheta \right)^n d\vartheta, \label{slsv1}
\end{eqnarray}
with the integration interval defined by (\ref{rect}) and
(\ref{curv}). 

\paragraph{${\cal P} \ge 0$}

In this case, the distribution function has to be integrated over the
grey shaded area in figure \ref{fam2int1}. Only if ${\cal P} < {\cal
A} b$ will the second of the curves (\ref{curv}) intersect the
rectangle (\ref{rect}).

\subparagraph{${\cal P} \ge {\cal A} b \ge 0$}

The $z'$-component of the angular momentum is positive everywhere
inside the rectangle (\ref{rect}). The integration over $\vartheta$
can be performed over the entire interval $[0,\pi]$, and so
\begin{eqnarray}
\int_0^\pi \left( {\cal P} + b v \cos
\vartheta \right)^n d\vartheta &=& \\
 &&\hspace{-6em}\pi {\cal
P}^n \,_2F_1 \left( -\frac{n}{2},\frac{1-n}{2};1;\left(\frac{bv}{{\cal
P}}\right)^2\right),
\end{eqnarray}
where $[n/2]$ is the largest integer, less or equal to $n/2$. Now, the
integration over $v$ can be tackled. This integral reduces to a
Beta-function and the spatial LOSVD can be
written in the following elegant form :
\begin{eqnarray}
\phi(\varpi,z',v_p) &=& 
\frac{2
\pi}{2^{n/2}} \frac{\varpi^n}{\alpha+1} \left( i_{\rm rot} b_1 v_p -
\frac{L_0}{\varpi} \right)^n \nonumber\\
&&  \hspace{-7em}\times\left( \psi-E_0-\frac{v_p^2}{2} \right)^{\alpha+1}  
 \nonumber \\ &&  \hspace{-7em}\times _2F_1 \left(
-\frac{n}{2},\frac{1-n}{2};\alpha+2; \frac{b^2 (2(\psi-E_0)-v_p^2)
}{(i_{\rm rot} b_1 v_p - \frac{L_0}{\varpi})^2} \right),
\label{slosvd1fam2}
\end{eqnarray}
which is finite and well-behaved for all integer $n$ and real
$\alpha$.

\subparagraph{${\cal P} \le {\cal A} b$}

In this case, $\cos \vartheta_0 \le 0$ and $\vartheta_0 \in
[\frac{\pi}{2},\pi]$. The second of the curves defined by (\ref{curv})
cuts out a piece of the rectangle (\ref{rect}) and hence the
distribution function has to be integrated over the shaded area in
figure
\ref{fam2int1} :
\begin{eqnarray}
\phi(\varpi,z',v_p) &=&  \frac{\varpi^n}{2^{\alpha+n/2}} 
\sum_{i=0}^n {n \choose i} {\cal P}^{n-i} b^i \nonumber\\
&&\hspace{-7em}\times
\left( \int_0^{\vartheta_0} (\cos \vartheta)^i \, d\vartheta 
\underbrace{\int_0^{{\cal A}^2} (v^2)^{i/2} ({\cal A}^2-v^2)^\alpha 
\, dv^2}_{\textstyle h_1} \right. 
\nonumber \\
&& \hspace{-7em}\left. + \int_{\vartheta_0}^\pi (\cos \vartheta)^i d\vartheta
\underbrace{\int_0^{(\frac{\cal P}{b \cos \vartheta})^2} (v^2)^{i/2} 
({\cal A}^2-v^2)^\alpha \, dv^2}_{\textstyle h_2}\right).
\end{eqnarray}
The first auxiliary function $h_1$ is simply a Beta-function
\begin{eqnarray}
h_1 &=& 
{\cal A}^{2(\alpha+1)+i} B(\alpha+1,\frac{i}{2}+1). \label{h1}
\end{eqnarray}
The second function $h_2$ is an incomplete Beta-function and can 
be rewritten in terms of a hypergeometric function :
\begin{eqnarray}
h_2 &=& 
\frac{2 {\cal A}^{2
\alpha}}{2+i} \left( \frac{\cal P}{b \cos \vartheta} \right)^{i+2} 
\nonumber\\&& \times\,_2F_1 \left( - \alpha, 1 + \frac{i}{2};2 + \frac{i}{2}; 
\left( \frac{\cal P}{{\cal A}b \cos \vartheta} \right)^2 \right). \label{h2}
\end{eqnarray}
Thus, we obtain the following expression for the spatial LOSVD
\begin{eqnarray}
\lefteqn{\phi(\varpi,z',v_p) = } \nonumber \\ && \hspace{-1.5em}\frac{\Gamma(\alpha+1)}{2^{\alpha+n/2}} \varpi^n 
{\cal A}^{2(\alpha+1)} {\cal P}^n\sum_{i=0}^n {n \choose i} 
\frac{\Gamma(1+\frac{i}{2})}{\Gamma(\alpha+2+\frac{i}{2})} 
\left( \frac{{\cal A}b}{\cal P} \right)^i\nonumber\\
&&\hspace{-1.5em}  \times\int_0^{\vartheta_0} (\cos
\vartheta)^i \, d\vartheta + 
 \frac{2 \varpi^n}{2^{\alpha+n/2}} {\cal A}^{2 \alpha}
\frac{{\cal P}^{n+2}}{b^2} \sum_{i=0}^n  \frac{1}{i+2} {n \choose i}  \nonumber \\
&& \hspace{-1.5em} \times\sum_{k=0}^\infty \frac{(-\alpha)_k (1+i/2)_k}{k!(2+i/2)_k} \left( \frac{\cal P}{{\cal
A} b} \right)^{2k} \int_{\vartheta_0}^\pi (\cos \vartheta)^{-2(k+1)} 
d\vartheta. \label{iets}
\end{eqnarray}
We first turn our attention to the second term of (\ref{iets}). Here,
only negative odd powers of $\cos \vartheta$ appear in the
integral. The integration can be done analytically with the aid of
Gradshteyn~\&~Ryzhik (1965) and yields the result :
\begin{eqnarray}
\int_{\vartheta_0}^\pi (\cos \vartheta)^{-2(k+1)} d\vartheta 
&=&\nonumber\\
&&\hspace{-6em}-\frac{\sin \vartheta_0}{2k+1} \sum_{j=0}^k
\frac{(-k)_j}{(\frac{1}{2}-k)_j} (\cos \vartheta_0)^{2(j-k)-1}. \label{thet1}
\end{eqnarray}
As to the first term of (\ref{iets}), the even and odd powers of $\cos
\vartheta$ have to treated separately
\begin{eqnarray}
\lefteqn{\sum_{i=0}^n {n \choose i} \frac{\Gamma(1+\frac{i}{2})}{\Gamma(\alpha+2+\frac{i}{2})} \left( \frac{{\cal A}b}{\cal P} \right)^i  
\int_0^{\vartheta_0} (\cos \vartheta)^i \, d\vartheta} \nonumber \\
&=& \frac{\theta_0}{\Gamma(\alpha+2)} \,_2F_1 \left(
\frac{n}{2},\frac{1-n}{2};\alpha+2; \left( \frac{{\cal A}b}{\cal P}
\right)^2 \right) \nonumber \\
&& \hspace{.5em} + \sin \vartheta_0 \sum_{j=0}^n \frac{(-n)_j}{\Gamma(j)} 
\frac{\Gamma(1 + \frac{j}{2})}{\Gamma(\alpha+\frac{j}{2}+2)} 
\left( -\frac{{\cal A}b}{\cal P} \right)^j \times \nonumber \\
&& \hspace{3.5em}\sum_{l=0}^{[\frac{j-1}{2}]}
\frac{((1-j)/2)_l}{(1-\frac{j}{2})_l} (\cos \vartheta_0)^{j-2l-1}. \label{thet2}
\end{eqnarray}
Thus, we are lead to the following expression for the spatial
LOSVD:
\begin{eqnarray}
\lefteqn{\phi(\varpi,z',v_p) = } \nonumber \\
&& \hspace{-2em}\frac{2 \vartheta_0}{2^{n/2}} \frac{\varpi^n}{\alpha+1} \left(
i_{\rm rot} b_1 v_p -\frac{L_0}{\varpi} \right)^n \left(
\psi-E_0-\frac{v_p^2}{2} \right)^{\alpha+1}
 \nonumber \\ &&  \hspace{-1em}\times _2F_1 \left(
-\frac{n}{2},\frac{1-n}{2};\alpha+2; \frac{b^2 (2(\psi-E_0)-v_p^2)
}{(i_{\rm rot} b_1 v_p - \frac{L_0}{\varpi})^2} \right) \nonumber \\
&& \hspace{-2em}+ 2 \Gamma(\alpha+1) \sin \vartheta_0 \frac{\varpi^n}{2^{n/2}} \left( \psi - E_0 -
\frac{v_p^2}{2} \right)^{\alpha+1} 
\left( i_{\rm rot}b_1 v_p - \frac{L_0}{\varpi} \right)^n  \nonumber \\
&& \hspace{-1em} \times\sum_{j=0}^n \frac{(-n)_j}{(j+1)!}
\left( - \frac{ {\cal A} b \cos \vartheta_0}{i_{\rm rot} b_1 v_p - \frac{L_0}{\varpi}} \right)^j\nonumber\\
 &&\hspace{-1em}\times
\sum_{l=0}^{[\frac{j-1}{2}]} \frac{((1-j)/2)_l}{(1-\frac{j}{2})_l} 
(\cos \vartheta_0)^{-2l-1} \nonumber \\
&& \hspace{-2em}- \tan \vartheta_0 \frac{\varpi^n}{2^{n/2}} \frac{1}{b^2} \left( \psi - E_0 -
\frac{v_p^2}{2} \right)^{\alpha} 
\left( i_{\rm rot}b_1 v_p - \frac{L_0}{\varpi} \right)^{n+2} \nonumber\\
&& \hspace{-1em}\times\sum_{i=0}^n \frac{1}{i+2} 
{n \choose i} \nonumber \\ &&  \hspace{-1em} \times\sum_{k=0}^\infty \frac{
(-\alpha)_k (1+\frac{i}{2})_k (\frac{1}{2})_k } {(\frac{3}{2})_k
(2+\frac{i}{2})_k k!}
\left( \frac{i_{\rm rot} b_1 v_p - \frac{L_0}{\varpi}}
{ b \cos \vartheta_0\sqrt{2(\psi-E_0)-v_p^2}} \right)^{2k} 
\nonumber \\ && \hspace{-1em} \times\,_2F_1 \left( -k,1;\frac{1}{2}-k; (\cos
\vartheta_0)^2 \right). \label{slosvd2fam2}
\end{eqnarray}
In the limit ${\cal P} \rightarrow {\cal A}b$, one finds that
$\vartheta_0 \rightarrow \pi$ and the complicated expression
(\ref{slosvd2fam2}) reduces to the far easier formula
(\ref{slosvd1fam2}), as it should. The rather daunting formula
(\ref{slosvd2fam2}) will only be well-behaved for integer values of
$\alpha$ since then the last summation (with summation variable $k$)
terminates after a finite number of terms.

\paragraph{${\cal P} < 0$}

\begin{figure}
\includegraphics[bb=50 55 545 730,angle=-90,scale=.4,clip=true]{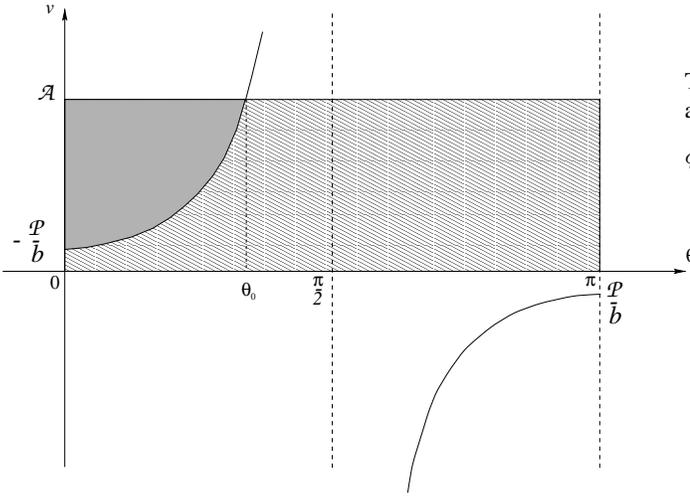}
\caption{The grey shaded area in the $(v,\vartheta)$-plane marks 
the area over which the distribution function has to be integrated to
calculate the spatial LOSVD in the case
${\cal P} < 0$. The hatched rectangle is defined by $0 \le v \le \cal
A$ and $0 \le \vartheta \le \pi$. The integration area is bounded by
the curve $v \ge \frac{-\cal P}{b \cos
\vartheta}$. \label{fam2int2}}
\end{figure}
In this case, $\cos \vartheta_0 = - \frac{\cal P}{{\cal A} b} > 0$ and
hence $\vartheta_0 \in [0,\frac{\pi}{2}]$. The distribution function
now has to be integrated over the shaded area in the
$(v,\vartheta)$-plane presented in figure \ref{fam2int2}. Only if
$|{\cal P}| < {\cal A} b$ will the first of the curves (\ref{curv})
intersect the rectangle (\ref{rect}).

\subparagraph{$|{\cal P}| \ge {\cal A} b$}

The integration area defined by the equations (\ref{curv}) and
(\ref{rect}) vanishes, hence the spatial LOSVD is identically zero for
all $|{\cal P}| \ge {\cal A} b$.

\subparagraph{$|{\cal P}| < {\cal A} b$}

The distribution function must be integrated over the grey shaded area 
in figure \ref{fam2int2} :
\begin{eqnarray}
\lefteqn{\phi(\varpi,z',v_p) =} \nonumber \\ &&  
\frac{\varpi^n}{2^{\alpha+n/2}} 
\sum_{i=0}^n {n \choose i} {\cal P}^{n-i} b^i 
\int_0^{\vartheta_0} (\cos \vartheta)^i \, d\vartheta  \nonumber\\
&&\times\int_{(\frac{|\cal P|}{b \cos \vartheta})^2}^{{\cal A}^2}
(v^2)^{i/2} ({\cal A}^2-v^2)^\alpha
\, dv^2.
\end{eqnarray}
We start with the integration over the velocity in the plane of the
sky, $v$, which can be performed analogously to (\ref{h1}) and
(\ref{h2}) :
\begin{eqnarray}
\lefteqn{\int_{(\frac{|\cal P|}{b \cos \vartheta})^2}^{{\cal A}^2}
(v^2)^{i/2} ({\cal A}^2-v^2)^\alpha \, dv^2} \nonumber \\ 
&=& {\cal
A}^{2(\alpha+1)+i} B(\alpha+1,\frac{i}{2}+1) \nonumber \\ 
&& - \frac{2}{i+2}
\left(\frac{|\cal P|}{b \cos \vartheta}\right)^{i+2} {\cal
A}^{2\alpha}\nonumber\\
&&\times
\,_2F_1\left(-\alpha,\frac{i}{2}+1;\frac{i}{2}+2;\left(\frac{|\cal
P|}{{\cal A} b \cos \vartheta}\right)^2\right).
\end{eqnarray}
Thus, we obtain an expression that is almost exactly like (\ref{iets})
\begin{eqnarray}
\lefteqn{\phi(\varpi,z',v_p) = } \nonumber \\&& \hspace{-2em}\frac{\Gamma(\alpha+1)}{2^{\alpha+n/2}} \varpi^n 
{\cal A}^{2(\alpha+1)} {\cal P}^n\nonumber\\
&&\hspace{-2em}\times\sum_{i=0}^n {n \choose i}
\frac{\Gamma(1+\frac{i}{2})}{\Gamma(\alpha+2+\frac{i}{2})} 
\left( \frac{{\cal A}b}{\cal P} \right)^i  \int_0^{\vartheta_0} (\cos
\vartheta)^i \, d\vartheta  \nonumber \\
&& \hspace{-2em}- \frac{2 \varpi^n}{2^{\alpha+n/2}} {\cal A}^{2 \alpha}
\frac{{\cal P}^{n+2}}{b^2} \sum_{i=0}^n  \frac{(-1)^i}{i+2} {n \choose i}  \nonumber \\
&&  \hspace{-2em}\times\sum_{k=0}^\infty \frac{(-\alpha)_k (1+i/2)_k}{k!(2+i/2)_k} \left( \frac{\cal P}{{\cal
A} b} \right)^{2k} \int_0^{\vartheta_0} (\cos \vartheta)^{-2(k+1)} 
d\vartheta. \label{iets2}
\end{eqnarray}
The integrations over $\vartheta$ are handled analogously to
(\ref{thet1}) and (\ref{thet2}) and we finally obtain
\begin{eqnarray}
\lefteqn{\phi(\varpi,z',v_p) = } \nonumber \\
&& \frac{2 \vartheta_0}{2^{n/2}} \frac{\varpi^n}{\alpha+1} \left(
i_{\rm rot} b_1 v_p -\frac{L_0}{\varpi} \right)^n \left(
\psi-E_0-\frac{v_p^2}{2} \right)^{\alpha+1}
 \nonumber \\ &&  \times_2F_1 \left(
-\frac{n}{2},\frac{1-n}{2};\alpha+2; \frac{b^2 (2(\psi-E_0)-v_p^2)
}{(i_{\rm rot} b_1 v_p - \frac{L_0}{\varpi})^2} \right) \nonumber \\
&& + 2 \Gamma(\alpha+1) \sin \vartheta_0 \frac{\varpi^n}{2^{n/2}} \left( \psi - E_0 -
\frac{v_p^2}{2} \right)^{\alpha+1} \nonumber\\
&&\times
\left( i_{\rm rot}b_1 v_p - \frac{L_0}{\varpi} \right)^n  \nonumber \\
&&  \times\sum_{j=0}^n \frac{(-n)_j}{(j+1)!}
\left( \frac{ {\cal A} b \cos \vartheta_0}{|i_{\rm rot} b_1 v_p - 
\frac{L_0}{\varpi}|} \right)^j \nonumber\\
&&\times
\sum_{l=0}^{[\frac{j-1}{2}]} \frac{((1-j)/2)_l}{(1-\frac{j}{2})_l} 
(\cos \vartheta_0)^{-2l-1} \nonumber \\ && - \tan \vartheta_0
\frac{\varpi^n}{2^{n/2}} \frac{1}{b^2} \left( \psi - E_0 -
\frac{v_p^2}{2} \right)^{\alpha} 
\left( i_{\rm rot}b_1 v_p - \frac{L_0}{\varpi} \right)^{n+2} \nonumber\\
&&\times\sum_{i=0}^n \frac{(-1)^i}{i+2} 
{n \choose i}  \nonumber \\ && \times  \sum_{k=0}^\infty \frac{
(-\alpha)_k (1+\frac{i}{2})_k (\frac{1}{2})_k } {(\frac{3}{2})_k
(2+\frac{i}{2})_k k!}
\left( \frac{i_{\rm rot} b_1 v_p - \frac{L_0}{\varpi}}
{ b \cos \vartheta_0\sqrt{2(\psi-E_0)-v_p^2}} \right)^{2k} 
\nonumber \\ &&  \times\,_2F_1 \left( -k,1;\frac{1}{2}-k; (\cos
\vartheta_0)^2 \right). \label{slosvd2fam22}
\end{eqnarray}
This formula will only be used for integer values of $\alpha$, as 
discussed in the case of (\ref{slosvd2fam2}). In the limit $|{\cal P}|
\rightarrow {\cal A}b$, $\vartheta_0$ vanishes. The spatial
LOSVD becomes identically zero and
remains zero for all $|{\cal P}| \ge {\cal A}b$.

\section[]{Dynamical model}
The two most prominent absorption lines of the Ca II triplet were used
as input for the modelling.  The spectra were rebinned to reach
sufficiently high S/N. The S/N was about 50 at radii less than 10~$''$ and from these spectra 182 points were included in the fit.  For
larger radii, the S/N was about 30 and only 91 points per spectrum
were included in the fit.

\subsection{Potential}\label{pot}

The observed R-band surface brightness profile was deprojected into a
spherical luminosity density. Assuming a mass-to-light ratio, this
lead to a mass density that was used to calculate a potential by means
of the Poisson equation. A model with constant M/L did not result in a
satisfactory fit. Hence, a number of models with a dark matter halo
were considered.  For all these models, the dark halo component was
added in the form of a spherical distribution of mass, where the
amount of matter is increasing as a power of the radius.  Different
models have different fractions of total matter mass to light emitting
mass at a chosen radius (here 2 kpc or $0.9 r_e$, this radius is only
for scaling and has no physical meaning) and different total
masses. In this way, a grid of models was created, and the smallest
value for $\chi^2$ was taken to be indicative for the best matching
potential.

\begin{figure}
\includegraphics[bb=30 80 440 574,clip=true,scale=.5]{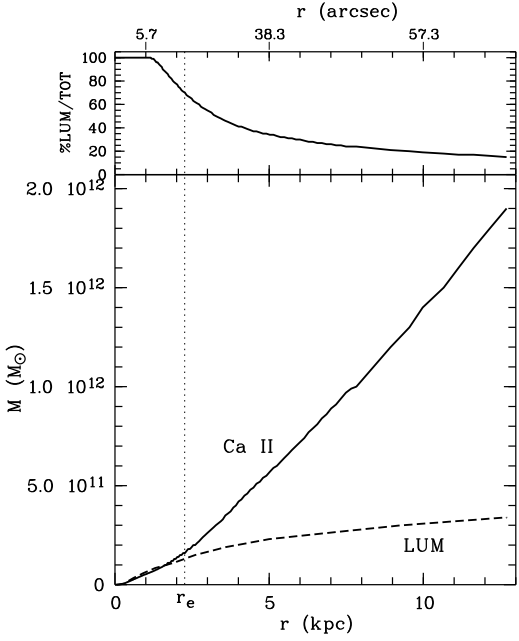}
  \caption{In the lower panel: Mass profile for the best fitting
  model based on the Ca II triplet feature (solid line) and for
  the best fitting model without dark matter (dashed line). In the
  upper panel: the fraction of luminous matter relative to all matter
  for the best fitting models with dark matter.}  \label{massar}
\end{figure}

The best results were obtained with a potential where the spatial
density of the total amount of matter was 2.5 times the spatial
density of the amount of luminous matter within 2 kpc. This scaling
implies that the total mass within $1 r_e$ is $1.6\times 10 ^{11}
M_\odot$, 70 $\%$ of which is luminous matter. The total mass at $2
r_e$ is $5\times10^{11} M_\odot$, and 37 $\%$ of this is luminous
matter (see also upper panel in figure \ref{massar}). Figure
\ref{massar} (lower panel) shows the total mass profile in solid line
in comparison with the luminous matter mass profile in dashed line.

It was not necessary to include a central black hole in the model to
obtain a good fit for the central spectra. 

\subsection{Dynamical model}
\begin{figure*}
\includegraphics[scale=.8]{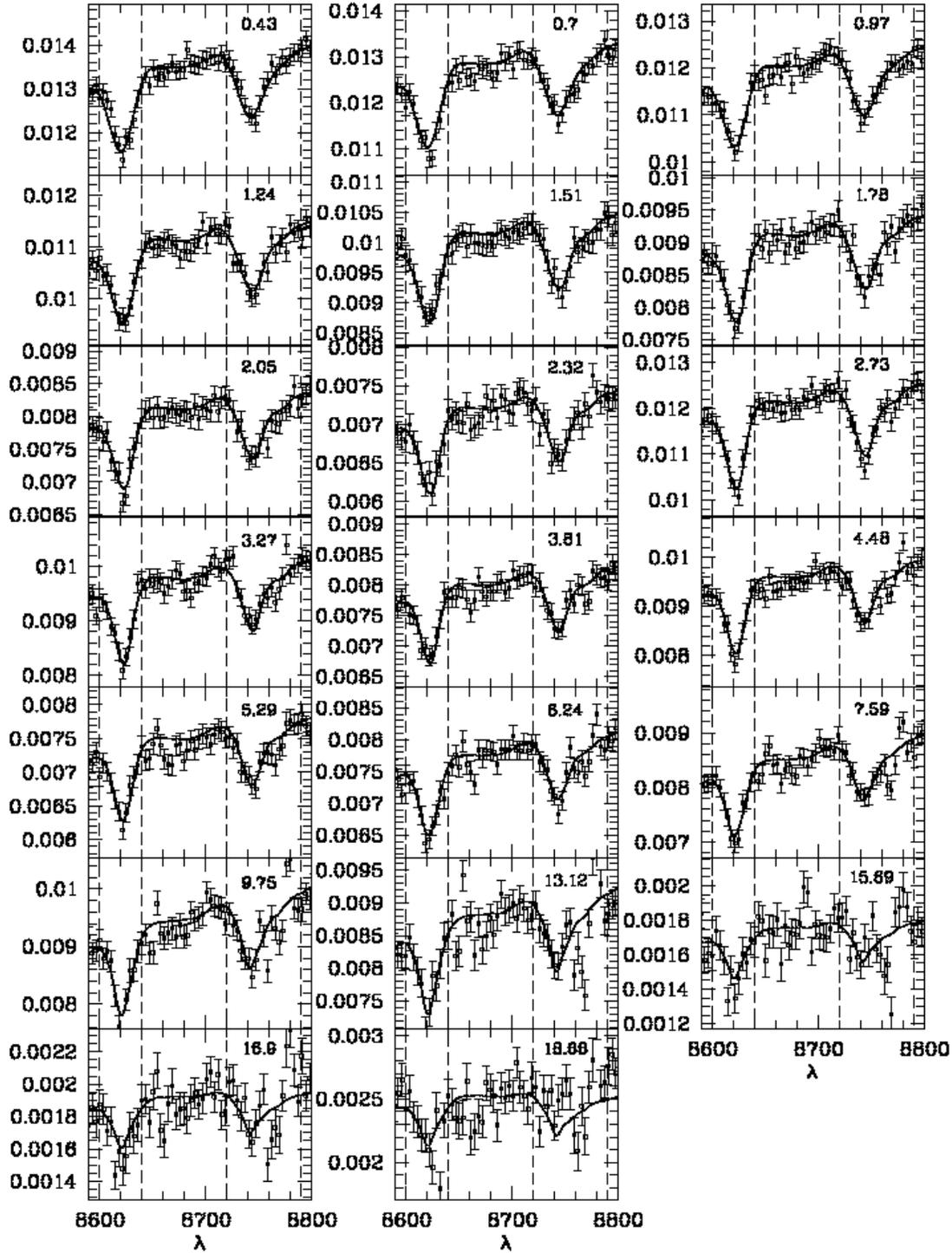} \caption{Data and model
  around the two strongest Ca II triplet lines. The
  regions included in the fit are located between the intervals
  indicated by dashed lines and with centre around 8620 {\AA} and 8750
  {\AA}. The projected radius of the positions where the spectra are
  taken are indicated in the panels.}  \label{specred}
\end{figure*}

The model that best fits the spectra around the two most prominent Ca
II triplet lines, at 8542 {\AA} and 8662 {\AA}, is shown in figure
\ref{specred}.  From these plots, no obvious shortcoming of the model
can be indicated.


\begin{figure}
\includegraphics[bb=40 430 460 690,scale=.6,clip=true]{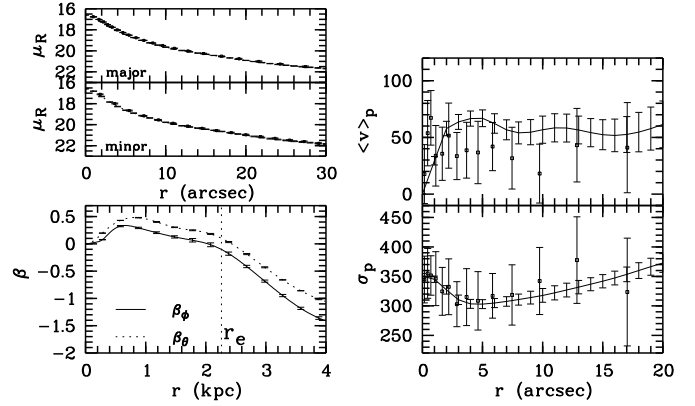}
  \caption{Some derived moments from the model. Left column: projected
  density on major and minor axis and the anisotropy parameters $\beta_\phi$ and
  $\beta_\theta$. Right column: projected mean velocity and projected velocity
  dispersion.}  \label{modred}
\end{figure}

Figure \ref{modred} shows some derived quantities of the model.  The
upper left panel shows the projected density on major and minor
axis. As can be seen, the flattening of the galaxy is well reproduced.
The right panels show calculated projected mean velocity and velocity
dispersion profiles (solid lines), together with uncertainties on these
quantities as derived from the model. The values and error estimates
for the projected mean velocity $\langle v_p \rangle$ and velocity dispersion $\sigma_p$, presented in \ref{data}, are plotted on top
in dots.  In figure \ref{modred} also the anisotropy parameters
$\beta_\phi = 1-(\sigma^2_\phi/\sigma^2_r)$ (in solid line) and
$\beta_\theta = 1-(\sigma^2_\theta/\sigma^2_r)$ (in dotted line) are
displayed in the lower left panel. The parameter $\beta_\phi$,
Binney's anisotropy parameter, is close to zero for the central parts,
indicating an isotropic DF there, becomes slightly posivite for
intermediate radii and slightly negative for radii larger than 2 kpc,
indicating that the model is tangentially anisotropic there.

\begin{figure}
\includegraphics[bb=55 315 255 400,scale=.8,clip=true]{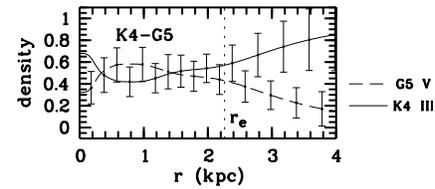} 
\caption{Template mixes for the best model based on the
two strongest lines of the Ca II triplet.}  \label{mixred}
\end{figure}

A mix of G5V and K4III stars gives the best fit to the spectra in this
wavelength range. The value for $\chi^2$ obtained for the fit is 2267,
the second best model has a $\chi^2$ of 2383. The resulting template
mix of this best model can be found figure \ref{mixred}.  The relative
densities are calculated from the spatial number densities of the
model.

More models, with other template mixes, than shown here were
calculated.  The results mainly confirm the trends seen in
figure \ref{mixred}.  In the model with K4III and G2V stars e.g., the
contribution of G2V stars to the density is larger than the G5V stars in
the K4III-G5V panel, but the same radial behaviour occurs.



\section{Discussion}

If a model is strongly tangential in the outer regions, most of the
motion occurs along the line of sight and this may increase the
projected velocity dispersion. Hence, the influence of dark
matter may be underestimated. This is the strongest argument for
including the fourth order shape parameter of the observed LOSVD as
input parameter in a dynamical model, because it is believed to
indicate whether a lot of orbits are observed near their tangent point
or not. For the calculation of this dynamical model, we did not rely
on the extracted LOSVD nor the derived kinematic parameters, but on
the observed spectra.

The dynamical model is slightly tangentially anisotropic from roughly
$1r_e$ on, as can be seen in figure \ref{modred}.

Tangential anisotropy in the outer regions of the model is also seen
in other models for e.g. NGC~3377 \citep{gebh}, NGC 1023 \citep{bow}
and also in the sample modeled by \citet{kron}.

Estimating kinematic parameters through a dynamical model clearly
reduces the amount of scatter in the kinematic profile, as is
illustrated in figure \ref{modred} (right panels).  This is easily
understood since the model induces additional dynamical constraints on
the parameters and as a result the values of the parameters taken at
different radii are not independent. When the parameters are inferred
directly from the observed spectra, there is no connection between the
parameters at different radii.

\begin{figure}
\includegraphics[bb=55 340 462 729,scale=.6,clip=true]{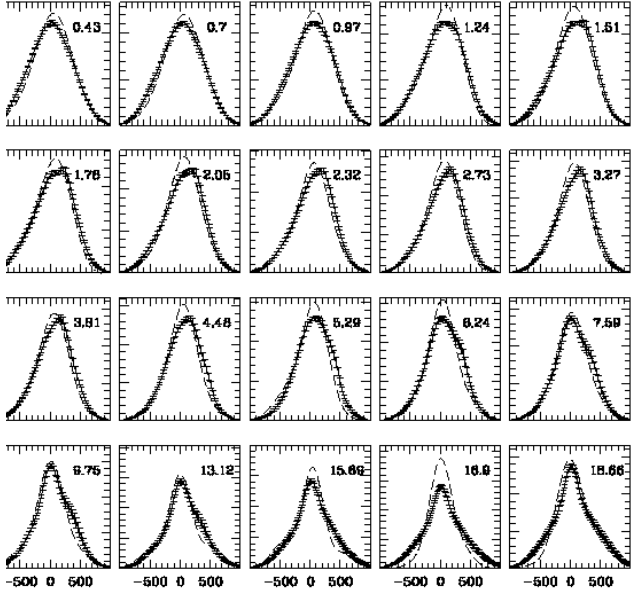} 
\caption{LOSVDs as reconstructed from the kinematic parameters
presented in \ref{data} (dashed line) and calculated from the model
(solid line). Also error bars are calculated from the model. The radii
at which the LOSVDs are taken are indicated in arcsec in the panels.}
\label{lpkes}
\end{figure}


Figure \ref{lpkes} presents LOSVDs as derived from the observations
and the model. The profiles calculated from the
kinematic parameters presented in \ref{data} are shown in dashed
lines, the solid lines present the LOSVDs with error bars calculated
from the model. It is clear that the LOSVDs calculated from the
kinematic parameters all have an approximately Gaussian shape. This is
of course due to the parametrization that is used to present them.
The LOSVDs calculated from the model show a large variety of
shapes. Some of these profiles cannot be reproduced by a Gauss-Hermite
series truncated at fourth order (or even sixth order) and hence a perfect
agreement with the LOSVDs calculated from the kinematic parameters cannot be expected. 

\begin{figure}
\includegraphics[bb=55 340 462 742,scale=.6,clip=true]{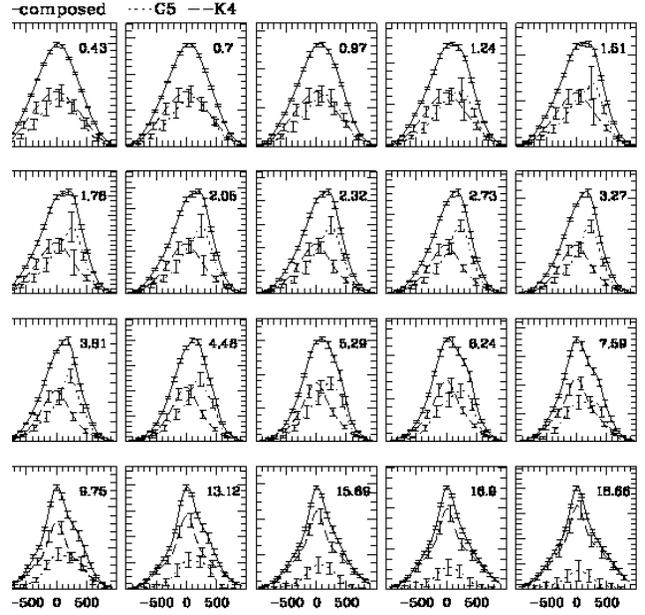} 
\caption{LOSVDs calculated from the model: for the composed DF in full
lines, for the G5V population in dotted lines and for the K4III population
in dashed lines. The different radii at which the LOSVDs are taken are
indicated in arcsec in the panels. }
\label{lppop}
\end{figure}

The shapes of the LOSVDs calculated from the model can be better
understood when looking at figure \ref{lppop}. In this figure the
LOSVDs for the different stellar populations, together with the
composed LOSVD (solid lines) are shown. The relative height of the
LOSVDs for the two populations is in agreement with the information in
figure \ref{mixred}. A full discussion of the role of the stellar
populations will be deferred to a subsequent paper, where a fit will
be presented that takes into account 2 wavelength regions.



\begin{figure}
\includegraphics[bb=40 80 465 445,scale=.6,clip=true]{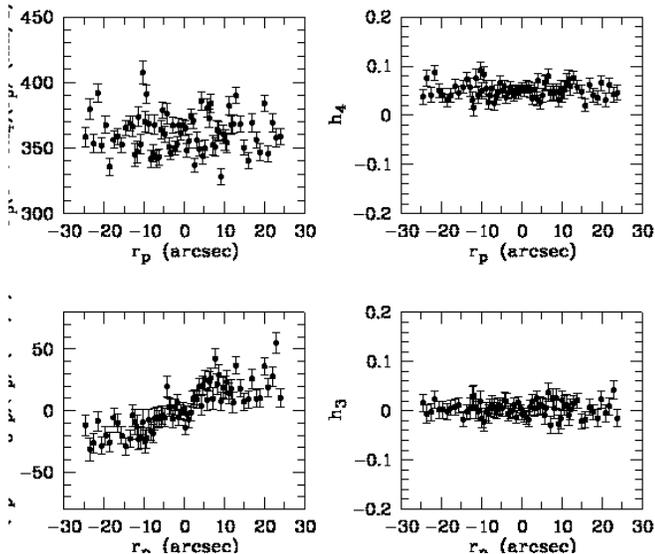}
\caption{Result from a simulation where $\langle v \rangle$, $\sigma$,
$h_3$ and $h_4$ are determined from spectra that are a convolution of
LOSVDs with $-20 \rm{km/s} < \langle v\rangle < 20 \rm{km/s}$, $\sigma
= 300 \rm{km/s}$, $h_3 = h_4 = h_5 = 0$ and $h_6 = 0.1$.}
\label{hasimul}
\end{figure}

Instead of comparing full profiles, one could argue in favor of
comparing their most relevant parameters.  Unfortunately, this is not
a trivial exercise. The mathematical elegane of the Hermite expansion
lies in the fact that it forms an orthogonal series, with coefficients
that can be calculated by integrals that follow from the
theory. However, this is almost never the way the Hermite coefficients
are calculated in practice: they are generally determined with some
kind of minimization after performing a convolution. This causes the 
loss of the orthogonality
property. This becomes clear in an exercise where a stellar template
spectrum convolved with a LOSVD with parameters $-20 \rm{km/s} <
\langle v \rangle < 20 \rm{km/s}$, $\sigma = 300 \rm{km/s}$ , $h_3 =
0$, $h_4 = 0$, $h_5 = 0$ and $h_6 = 0.1$ is analysed in the same way
as real galactic spectra and kinematic parameters up to $h_4$ are
determined. Since the template spectrum used to create this synthetic
spectrum is equal to the one used to determine the parameters again,
there is no influence of continuum subtraction or template mismatch on
the results. The synthetic galaxy spectrum has a S/N of 130 in the
centre and the light profile follows a de Vaucouleurs that resembles
the light profile of NGC~3258.  The result of this simulation is shown
in figure \ref{hasimul}. The values retrieved for $\langle v \rangle$
and $h_3$ correspond to the original values. On the other hand, the
values for $\sigma$ and $h_4$ are clearly higher than the values of
the input LOSVDs.

This indicates that the kinematic parameters as derived from a
Gauss-Hermite parametrization are in practice not independent, in the
sense that power in the higher orders than considered in the expansion
may contaminate lower order parameters. In the case of $h_4$ this can
give rise to wrong conclusions concerning the anisotropy in the
stellar motion.  Naively, one could conclude that it is better to go
to higher order than $h_4$ in the parametrization. Unfortunately, the
signal-to-noise ratio in outer regions of a galaxy is hardly
sufficient to derive reliable values for $h_4$, so it does not make
sense to try to determine higher order parameters.  The useful
information in LOSVDs derived from observations is thus very limited,
due to the rather low signal-to-noise of the observations. Therefore,
though it is a good idea to use a parametrization with a limited
number of parameters to represent these LOSVD, there is, as a
consequence, also only a limited number of shapes for LOSVDs
possible, as extracted from these observations. This on top of the
fact that these parameters are not independent, which means that
caution should be taken when interpreting them.  On the other hand,
LOSVDs calculated from dynamical models can have a much larger variety
of shapes. It would be unfair to reject model LOSVDs because they do
not match the ones that are derived from observations, simply because
the information in LOSVDs derived from spectra is very limited, while
the dynamical models offer much more freedom.

\section{Conclusions}

In this paper we use a method to analyse spectra from elliptical
galaxies in order to retrieve dynamical information and to perform a
stellar population synthesis at the same time. The method originates
from the field of dynamical modelling and is presented by DD98.  The
main idea is that the different stellar populations that contribute to
the integrated galaxy spectrum do not necessarily share the same
kinematic characteristics. Hence, in terms of dynamical modelling,
they should be given different distribution functions. This paper
reports for the first time on the application of our modelling
technique to data. The attention is focused on the presentation of the
method and on establishing a mass estimate for the galaxy under
study. A second paper will discuss the possibilities for stellar
population studies and will go deeper into the distribution functions
for separate stellar templates.

Modelling the spectra directly yields an estimate for the total amount
of mass, and an analysis of the internal dynamics of the galaxy.  This
technique, using galaxy spectra as input data, also differs from more
traditional modelling strategies that rely on kinematic parameters
that are derived from the spectra in the sense that a two step process
is turned into a one step process, thereby coming closer to the
astrophysical issues.

Photometric and spectroscopic observations for NGC~3258 are
presented. This galaxy has almost perfect elliptical isophotes, and
has an effective radius of $12.95 ''$ (2.26 kpc). Images in B and R show
a slightly off-centre nucleus. There seems to be a slight reddening of
the nucleus. The photometry also reveals the presence of a central
dust disk (0.17 kpc - 0.26 kpc). The R band photometry was used to
derive a spherical potential for the luminous matter.

The photometry, together with galaxy spectra containing dynamical
information up to $1.3 r_e$ are combined into a dynamical model.  The
best dynamical model for NGC~3258 is the one with a dark matter halo
that contributes 30\% of the total mass at $1 r_e$ and 63\% of the
total mass at $2 r_e$. From $1.45 r_e$ outwards, the model is
dominated by dark matter. The total mass within $1 r_e$ is $1.6\times
10 ^{11} M_\odot$ and $(M/L)_R \approx 65$, $(M/L)_R \approx 30$,
whereas the total mass extrapolating to $2 r_e$ is $5\times10^{11}
M_\odot$ and $(M/L)_R \approx 95$,$(M/L)_B \approx 43.5$. In the
region without dark matter $(M/L)_R \approx 14$ ($(M/L)_B \approx
6.4$). The modelling did not require to include a central black hole.

The anisotropy parameter (see figure \ref{modred}) indicates that the
model is isotropic in the centre, is radially anisotropic between 0.2
and 2 kpc (i.e. within $1 r_e$) and becomes slightly tangentially
anisotropic further out.

A simulation has shown that for the analysis of LOSVDs derived from
observations by means of a truncated Gauss-Hermite series, power in
higher orders than considered in the expansion may contaminate lower
order parameters. This is a worrysome result, since $h_4$ profiles
derived from observations tend to be interpreted in terms of
anisotropy of the galaxy. If this parameter is contaminated by power
in higher orders by some form of aliasing induced by the fitting
process, this may lead to an erroneous interpretation or at least may
make comparison of $h_4$'s obtained by different methods problematic.
Moreover, due to the freedom in the dynamical model, one cannot expect
that LOSVDs calculated from a dynamical model as presented in this
paper can be approximated by Gauss-Hermite series that are truncated
at a low order.

\section{Acknowledgments}
VDB acknowledges financial support from FWO-Vlaanderen.  WWZ
acknowledges the support of the Austrian Science Fund (project P14783)
and the support of the Bundesministerium f\"ur Bildung, Wissenschaft
und Kultur.

This research has made use of the NASA/IPAC Extragalactic Database
(NED) which is operated by the Jet Propulsion Laboratory, California
Institute of Technology, under contract with the National Aeronautics
and Space Administration.

\bsp \label{lastpage} 
\begin{thebibliography}{}
\bibitem[Bettoni \& Buson, 1987]{ber} Bettoni D., Buson L.M., 1987, A\&AS, 67, 341
\bibitem[\protect\citeauthoryear{Bower et al}{2001}]{bow} Bower G.A., Green R.F., Bender R., Gebhardt  K., Lauer T.R., Magorrian J., Richstone D.O., Danks A., Gull T., Hutchings J., Joseph C., Kaiser M. E., Weistrop D., Woodgate B., Nelson C., Malumuth E.M., 2001, ApJ, 550, 75.
\bibitem[\protect\citeauthoryear{Bregman et al.}{1998}]{breg} Bregman J.N., Snider B.A., Grego L., Cox C.V., 1998, ApJ, 499, 670
\bibitem[Carter et al., 1999]{cbh} Carter D., Bridges T.J., Hau G.K.T., 1999, MNRAS, 307, 131
\bibitem[Ciotti et al., 1995]{ci} Ciotti et al., 1995, MNRAS, 276, 961 
\bibitem[De Bruyne et al., 2001]{db} De Bruyne V., Dejonghe H., Pizzella A., Bernardi M., Zeilinger W.W., 2001, ApJ, 546, 903
\bibitem[De Bruyne et al., 2003]{db2} De Bruyne V., Vauterin P., De Rijcke S., Dejonghe H., 2003, MNRAS, 339, 215
\bibitem[Dejonghe, 1989]{dj} Dejonghe H., 1989, ApJ, 343, 113
\bibitem[Dejonghe \& De Bruyne, 2003]{job} Dejonghe H., De Bruyne V.,
'From observations to distribution function', in Lecture Notes in
Physics 'Galaxies and Chaos. Theory and Observations', Springer Verlag
\bibitem[\protect\citeauthoryear{De Rijcke \& Dejonghe}{1998}]{dr} De Rijcke S., Dejonghe H., 1989, MNRAS, 298, 677 (DD98)
\bibitem[\protect\citeauthoryear{De Rijcke et al.}{2003}]{rijdi} De Rijcke S., Dejonghe H., Zeilinger W.W., Hau G.K.T., 2003, A\&A, 400, 119 
\bibitem[\protect\citeauthoryear{Fabiano et al.}{1992}]{fab} Fabbiano G., Kim D.-W., Trinchieri G., 1992, ApJS, 80, 531
\bibitem[\protect\citeauthoryear{Ferrarese \& Merritt}{2000}]{ferm} Ferrarese L., Merritt D., 2000, ApJ, 539, L9
\bibitem[\protect\citeauthoryear{Ferrari et al.}{2002}]{fer} Ferrari F., Pastoriza M.G., Macchetto F.D., Bonatto C., Panagia N., Sparks W.B., 2002, A\&A, 389, 355
\bibitem[Fisher et al., 1995]{fib} Fisher D., Franx M., Illingworth G., 1995, ApJ, 448, 119

\bibitem[\protect\citeauthoryear{Gebhardt et al.}{2000}]{gebh} Gebhardt K., Richstone D., Kormendy J., Lauer T. R., Ajhar E.A., Bender R., Dressler A., Faber S. M., Grillmair C., Magorrian J., Tremaine S., 2000, AJ, 119, 1157
\bibitem[\protect\citeauthoryear{Gerhard et al.}{1998}]{gerh} Gerhard O., Jeske G., Saglia R.P., Bender R., 1998, MNRAS, 295, 197
\bibitem[\protect\citeauthoryear{Gradshteyn \& Ryzhik}{1965}]{gr}Gradshteyn I.S., Ryzhik I.M., Table of integrals, series and products, 1965, Academic Press, New York\&London
\bibitem[\protect\citeauthoryear{Jenkins}{1983}]{jen} Jenkins C.R., 1983, MNRAS, 205, 1321 
\bibitem[\protect\citeauthoryear{Joseph et al.}{2001}]{jos} Joseph, C.L., Merritt D., Olling R., Valluri M., Bender R., Bower G., Danks A., Gull T., Hutchiongs J., Kaiser M.E., Maran S., Weistrop D., Woodgate B., Malumuth E., Nelson C., Plait P., Lindler D., 2001, ApJ, 550, 668
\bibitem[\protect\citeauthoryear{Koprolin \& Zeilinger}{2000}]{kop} Koprolin W., Zeilinger W.W., 2000, A\&AS, 145, 71
\bibitem[\protect\citeauthoryear{Kronawitter et al.}{2000}]{kron} Kronawitter A., Saglia R.P., Gerhard O., Bender R., 2000, A\&AS, 144, 53
\bibitem[\protect\citeauthoryear{Pellegrini et al.}{1997}]{pel} Pellegrini S., Held E.V., Ciotti L., 1997, MNRAS, 288, 1
\bibitem[\protect\citeauthoryear{Sadler et al.}{1989}]{sad} Sadler E.M., Jenkins C. R., Kotanyi C.G., 1989, MNRAS, 240, 591 
\bibitem[\protect\citeauthoryear{van der Marel}{1994}]{vdm} van der Marel R. P., 1994, MNRAS, 270, 271
\bibitem[\protect\citeauthoryear{van der Marel \& Franx}{1993}]{vdmf} van der Marel R. P., Franx M., 1993, ApJ, 407, 525
\end{thebibliography}
\end{document}